\documentclass[twocolumn,showpacs,preprintnumbers,amsmath,amssymb,
showkeys,floatfix,aps,superscriptaddress]{revtex4}
\usepackage{graphicx}
\usepackage{dcolumn}
\usepackage{bm}
\usepackage{bigints}
\usepackage{natbib}
\usepackage{amsmath}
\begin{document}
{ \footnotesize \quad Phys. Rev. D 109, 023523 \quad Published 19 January 2024  \qquad \href{https://link.aps.org/doi/10.1103/PhysRevD.107.063506}{DOI: 10.1103/PhysRevD.109.023523}}
\title{ Least response method to separate CMB spectral distortions from
  foregrounds}
\author{J.-P. Maillard}
\affiliation{Institut d'Astrophysique de Paris, UMR 7095, CNRS, Sorbonne Université, 98bis Boulevard Arago, 75014 Paris, France}
\author{A. Mihalchenko}
\affiliation{Astro-Space Center of P.N. Lebedev Physical Institute, Profsoyusnaya 84/32, Moscow 117997, Russia}
\affiliation{Moscow Institute of Physics and Technology, Institutskiy pereulok, d.9, Dolgoprudny, Moscow 141701, Russia}
\author{D. Novikov}
\affiliation{Astro-Space Center of P.N. Lebedev Physical Institute, Profsoyusnaya 84/32, Moscow 117997, Russia}
\author{A. Osipova}
\affiliation{Astro-Space Center of P.N. Lebedev Physical Institute, Profsoyusnaya 84/32, Moscow 117997, Russia}
\affiliation{National Research University Higher School of Economics, 11 Pokrovsky Bulvar,
Moscow 109028, Russia
}
\author{S. Pilipenko}
\affiliation{Astro-Space Center of P.N. Lebedev Physical Institute, Profsoyusnaya 84/32, Moscow 117997, Russia}
\author{J. Silk}
\affiliation{Institut d'Astrophysique de Paris, UMR 7095, CNRS, Sorbonne Université, 98bis Boulevard Arago, 75014 Paris, France}
\affiliation{Department of Physics and Astronomy, The Johns Hopkins University, Baltimore, Maryland 21218, USA}
\affiliation{Beecroft Institute for Particle Astrophysics and Cosmology, Department of Physics, University of Oxford, Oxford OX1 3RH, United Kingdom}

\begin{abstract}
  We present a signal-foreground separation algorithm for filtering
  observational data to extract spectral distortions of the cosmic microwave
  background (CMB).
  Our linear method, called the least response method (LRM), is based on the
  idea of simultaneously minimizing the response to all possible foregrounds
  with poorly defined spectral shapes and random noise while maintaining a
  constant response to the signal of interest. This idea was introduced in
  detail in our previous paper. Here, we
  have expanded our analysis by taking into consideration all the main
  foregrounds. We draw a detailed comparison between our approach and the
  moment internal linear combination method, which is a modification of
  the internal linear combination technique previously used for CMB anisotropy maps. We
  demonstrate advantages of LRM and evaluate the prospects for measuring various
  types of spectral distortions. 
  
  Besides, we show that LRM suggests the possibility of its
  improvements if we use an iterative approach with sequential separation
  and partial subtraction of foreground components from the observed
  signal.
  
  In addition, we estimate the optimal
  temperature that the telescope's optical system should have in order to detect
  the chemical type $\mu$ distortions. We present a design of an instrument
  where, according to our estimates, the optimal contrast between its thermal
  emission and the CMB allows us to measure such distortions.
\end{abstract}


\maketitle

\section{Introduction}
The recent space missions related to the study of relic radiation WMAP
\citep{2013ApJS..208...20B} and Planck
\citep{2014A&A...571A..15P,2018arXiv180706208P} has substantially broadened our
knowledge of cosmic microwave  background anisotropy: the power spectrum of
$\Delta T/T$ fluctuations and almost Gaussian nature of their
distribution on the celestial sphere. Ultimately, these observational data
helped to estimate the main cosmological parameters.

However, even more information can be gained
by delving into cosmic microwave background (CMB) spectral distortions (SDs). Being one of the key goals of
observational cosmology \cite{2021ExA....51.1515C, 2021APS..APRB21003S, 2012MNRAS.419.1294C, 2014Sci...344..586S, 2016JCAP...03..047D, 2016MNRAS.460..227C,  2014PTEP.2014fB107T, 2016PhRvD..93h3515C}, measuring
deviations of the CMB spectrum from
a black body shape shall reveal a
vast amount of information about the early Universe, unobtainable by other
observational methods
\citep{1969Ap&SS...4..301Z, 1991A&A...246...49B, 2019BAAS...51c.184C, 2021ExA....51.1515C,  2018PhRvD..97d3525N}.

The chemical potential $\mu$-type distortions \citep {1970Ap&SS...7...20S} are created if
either energy is injected into
the cosmic plasma or photon number density is modified \citep{2021JCAP...12..050F, 1970Ap&SS...9..368S,1991ApJ...371...14D, 1994ApJ...430L...5H, 2012MNRAS.419.1294C, 2014JCAP...10..029O, 2022MNRAS.515.5775A} when the redshift is less than $\sim 2\times 10^6$. Thus, they can be used as a tool to
trace the energy history of the Universe.
A later type of distortions ($y$ distortions)  \citep{1969Ap&SS...4..301Z} contain information about the structure of
intracluster medium \citep{2011A&A...535A.108C, 2011A&A...527L...1C, 2000A&A...360..417E, 2015MNRAS.452.1328M}. In addition, very specific
deviations in the frequency spectrum of radiation coming from galaxy clusters
can be used for independent measurements of the amplitudes and
orientations of low CMB anisotropy multipoles \citep{2018PhRvD..98l3513E, 2020PhRvD.101l3510N}. The COBE/FIRAS mission \citep{1990ApJ...354L..37M,Fixsen_2009}
measured the CMB frequency spectrum and established its Planckian shape with
good accuracy. Future detection of the spectral features is expected from 
FIRAS-type missions \citep{A.Kogut_2011, 2016SPIE.9904E..0WK, 2017MNRAS.471.1126A, 2023JCAP...07..057K}, experiments with a large primary mirror with good angular resolution \citep{2021PhyU...64..386N} (which is essential for $y$-type deviations) or even using a
Moon-based telescope \citep{doi:10.1098/rsta.2020.0212}, which is revised
in this paper.

An exhaustive review of the theory behind SDs, their observational challenges,
computational techniques and possible new directions in this field are given
in \citep{2021JCAP...12..050F}. The main difficulty in the task of measuring
small
spectral distortions is the presence of foregrounds of cosmic and instrumental
origin, which are several orders of magnitude larger in amplitude than the
signals of interest. Thus, in order to ensure detection of such
signals, it is necessary not only to achieve high sensitivity of the
experiment, but also to learn the most optimal way to separate SDs from the rest of
the observed signal.
The spectra of some astronomical foregrounds as well as the spectrum emitted
by the optical system of the instrument are poorly defined and cannot be modeled
and predicted with the accuracy required for
reliable SDs measurements. This means that sufficiently advanced and efficient
data processing methods must be used to solve such a problem. In particular,
any methods used must take into account possible variations in foreground
frequency spectra.

The well-known blind internal linear combination (ILC) method \citep{1992ApJ...398..169R} has been
successfully applied to process
multifrequency analysis of CMB anisotropy maps \citep{2003PhRvD..68l3523T, 2003ApJS..148...97B, 2004ApJ...612..633E}. However, the application of this
method is limited by the presence of large foreground components that have a
nonzero projection onto the signal of interest, which leads to a bias.
To avoid this, the constrained ILC (cILC) \citep{2011MNRAS.418..467R, 2020MNRAS.494.5734R, Hill:2013baa} method was proposed as a modification of the ILC.
This approach completely eliminates the contribution of foregrounds with
well-known spectral shapes, treating the remaining components as unmodeled noise.
However, to solve the problem of getting rid of all foreground components,
it is necessary to take into account the spectral variations of dust, cosmic infrared background (CIB), synchrotron radiation and other sky foregrounds along the line of site and from one direction to another one, as well as time variations of the foreground contribution from the instrument emission. The spectra of these components depend on the
parameters and, therefore, spectral variations are equivalent to variations in
the parameters.

To take this into account, a quite effective method,
called MILC (Moments ILC) 
\citep{2005MNRAS.357..145S, 10.1093/mnras/stx1982, 10.1093/mnras/staa3292}, was introduced, where the foreground spectra are expanded into Taylor series
in terms of parameters in the vicinity of some average reference parameter values.
Data filtering, which involves zeroing out the expansion moments, ensures that
the contribution from such foregrounds is eliminated. However, this method also
has a rather serious disadvantage. A large number of constraints ensuring the
zeroing of the foreground contribution during the filtering process inevitably
leads to an unacceptably large response to unmodeled noise (including random
photon noise). This is not surprising, since a large number of imposed strict
conditions limits the number of degrees of freedom and increases the
contribution of noise to the estimation of the signal of interest.

In work \citep{Abylkairov_2021}, another modification of the cILC
method, called partially constrained ILC (pcILC), was proposed. Instead of
completely eliminating the contribution from a foreground with a well-defined
spectrum (as cILC does), pcILC reduces this contribution to some empirically
determinable level. As a result the strict cILC constraints are
somewhat softened. This made it possible to reduce the response to noise
when processing $\Delta T/T$ anisotropy data in the presence of the
Sunyaev-Zel'dovich effect or CMB lensing. However, application of this approach
in the presence of many different foregrounds is very complicated and its
effectiveness in this case has not been proven. Besides, unlike MILC, this
method does not allow for possible spectral shape variations of foregrounds.

We recently proposed a data filtering method called least response method (LRM)
\citep{2023PhRvD.107f3506N}, which is quite simple and easy to implement. This
approach involves the optimization of a single functional for all components of
the observed signal. During the process of data filtering we minimize the response to all
foregrounds and random noise simultaneously, while keeping the response to the
signal of interest equal to unity. We assume that we know the following information
about foreground signals:

1. The foreground spectral parameters can change within a limited range of their possible variations;

2. The amplitudes of the foregrounds are limited from above by known values.

Note that such information is available to us from previous observational data and
characteristics of the instrument optical system for a particular experiment.
In this article, we compare LRM and MILC
approaches and show the advantages and prospects of LRM for measuring various
spectral distortions of the relic radiation in the presence of all main foregrounds,
including the signal from the instrument optics.

In addition, we show that when measuring $\mu$ distortions, the temperature
of the instrument optical system should not approach the CMB temperature.
Measuring such distortions requires calibration of the instrument and
in this case the signal component generated by the optics becomes a part
of the observed signal. This component is close in shape to a black body
signal, and if its temperature is close to 2.7 K, then it becomes
'poisoned' to some extent, since the device itself begins to create spectral
features simulating CMB distortions. We show that, according to our estimates,
the optimal temperature of the instrument for $\mu$ spectral distortion
measurements should be close to 9 K.

The  outline of this paper is as follows.
In Sec. II, we review linear methods for observational data filtering.
We describe MILC and LRM approaches for CMB spectral distortions
detection in the case when the observed signal contains foregrounds with
poorly defined spectral shapes. In Sec. III, we describe models of
signals of interest, foregrounds and photon noise and present a detailed
comparative numerical analysis of MILC and LRM methods. In this Section
we also estimate the instrument optics temperature that is optimal for
detecting $\mu$-type CMB spectral distortions. In the same section we give
an example of an instrument for measuring the  relic radiation spectral
features.
Brief conclusions and our suggestions about possible improvements of
the LRM approach are given in Sec. IV.

\section{Linear Data Filtering Methods}

The frequency spectrum $S(\nu)$ we observe consists of
the signal of interest and a set of other components that we would like to
get rid of during data processing. Therefore, the total observed spectrum can
be written as follows:
\begin{equation}
    S(\nu)=a_d I_d(\nu)+\sum\limits_{m=1}^MI_m(\nu),
\end{equation}
where $I_d$ is the certain kind of CMB spectral distortions,
which we would like to separate from other components. Index ``$d$'' can
denote the $\mu$ distortions: $I_\mu$, the Sunyaev-Zel'dovich effect
($y$ distortions): $I_{y_0}$, the first or second relativistic corrections to this
effect: $I_{y_1}$ or $I_{y_2}$. Therefore, $a_d$ is the frequency-independent
amplitude of the signal
of interest we want to estimate. The rest of the signal $S(\nu)$
consists of $M$ foregrounds of various physical origins $I_m(\nu)$,
which may include spectral distortions themselves. For example, if we
are interested in signal $I_\mu$, then signals $I_{y_0},I_{y_1}I_{y_2}$ will
be part of the total foreground.

Using a Fourier-transform spectrometer (FTS)
mounted on a space telescope, one can obtain discrete values of the total signal
(or row vector) ${\bf S}=(S_1,.., S_J)$ in $J$ equally wide frequency channels
in a broad frequency range from $\nu_{min}$ to $\nu_{max}$:
\begin{equation}
  \begin{array}{l}
    \vspace{0.2cm}
  S_j=a_dI_d^j+\sum\limits_mI_m^j+N_j,\hspace{0.2cm}j=1,..,J,\\
   I_x^j=\int\limits_{{}_{\nu_j-\frac{\Delta\nu}{2}}}^{{}^{\nu_j+\frac{\Delta\nu}{2}}}
   \hspace{-0.2cm}I_x(\nu)\frac{d\nu}{\Delta\nu},
   \end{array}
\end{equation}
where indices $j$ indicate the frequency channel number, $\Delta\nu$ is the channel width and $N_j$ is the random photon noise with
zero mean and covariance matrix
$\left[C_{ij}\right]={\bf C}=\langle{\bf N^T}{\bf N}\rangle$.
The spectral shapes of
CMB distortions $I_\mu$, $I_y$, $I_{y1}$ and $I_{y2}$ are well-known,
while the foreground spectra $I_m$ can depend on various parameters and in
fact are
superpositions of spectra integrated along the line of sight or/and
obtained as a result of averaging over spatial pixels of a sky map.
Therefore, the foreground components can be written as
\begin{equation}
  \begin{array}{l}
    \vspace{0.2cm}
    I_m^j=\int\limits_{\Omega} a_m({\bf P})
    f_m(\nu_j,{\bf P})d{\bf P},\\
    d{\bf P}=dp_{{}_1}dp_{{}_2}\cdot\cdot dp_{{}_L},
 \end{array}   
\end{equation}
where ${\bf P}=p_{{}_1},..,p_{{}_L}$ is the set of $L$ parameters,
$f_m(\nu_j,{\bf P})$ are the functions representing the foreground spectra,
$\Omega$ is the region of possible parameter variations, and $a_m$ are the
amplitudes of the foreground radiation as functions of
parameters ${\bf P}$. Thus, if, for example, $a_m({\bf P})$ has the form of
a delta function $a_m({\bf P})=A_m\cdot\delta({\bf P}-{\bf P_m})$, then the
foreground spectrum with index $m$ will have a
template with well-defined parameters ${\bf P_m}$ and the amplitude $A_m$:
$I_m^j=A_m\cdot f_m(\nu^j,{\bf P_m})$.

The total observed signal ${\bf S}=(S_{{}_1},..,S_{{}_J})$ 
can be divided into three parts (three vectors):
\begin{equation}
  \begin{array}{l}
    \vspace{0.2cm}
   {\bf S}=a_d{\bf I_{d}}+{\bf F}+{\bf {N}},\\
   {\bf F}=(F_1,..,F_J),\hspace{0.2cm}F_j=\sum\limits_mI_m^j,\\
   {\bf N}=(N_1,..,N_J)
   \end{array}
\end{equation}
where ${\bf I_d}$ is some particular kind of spectral distortion
we want to separate from the rest of the signal, ${\bf F}$ is the total
foreground and ${\bf N}$ represents the random noise.

The task of any linear algorithm is to find the  optimal vector of weights
$\boldsymbol\omega=(\omega_1,..,\omega_J)$
for frequency channels that should have the following property:
\begin{equation}
  \boldsymbol{\omega}{\bf S^T}=\sum\limits_{j=1}^{J}\omega_jS_j
  \rightarrow a_d\hspace{0.1cm}
  for\hspace{0.1cm}C_{ij}\rightarrow 0,\hspace{0.2cm}i,j=1,..,J.
\end{equation}
Thus, in the ideal case, when the noise tends to zero
($C_{ij}\rightarrow 0$), it is desirable that
the algorithm accurately reproduces the amplitude of the signal of interest
$a_d$ without any bias. With nonzero noise, the algorithm should minimize the
deviation of the estimate of this amplitude from its true value.

Let us denote the scalar product $\boldsymbol{\omega}{\bf S^T}=R({\bf S})$ as
the response to the signal:
\begin{equation}
    R({\bf S})=a_d R({\bf I_d})+R({\bf F})+R({\bf N}).
 \end{equation}
The first condition imposed on the weights is common to all linear algorithms
and is quite obvious:
\begin{equation}
R({\bf I_d})=\boldsymbol{\omega}{\bf I_d^T}=\sum\limits_j\omega_jI_d^j=1.
\end{equation}
Thus, it is necessary to minimize the response to the rest of the signal
$R({\bf F})+R({\bf N})$ while maintaining the condition in Eq. 7.
To denote the expected total average value of the response to
foreground and noise, we use the following definition:
\begin{equation}
  R({\bf F}+{\bf N}):=\sqrt{R^{{}^2}({\bf F})+\langle R^{{}^2}({\bf N})\rangle}.
\end{equation}
Therefore, $R({\bf F}+{\bf N})$ is the value that must be minimized under
condition in Eq. (7).

\subsection{Internal linear combination method}

The ILC method assumes that we know the 
spectral energy distribution (SED)
of the signal of interest, but does not assume knowledge of any information
about foregrounds and noise. With this approach all components of the
observed signal, apart from the signal of interest, are considered as
unmodeled noise. That is, in this case in Eqs. (4) and (8) no distinction is made
between the components ${\bf F}$ and ${\bf N}$. Thus, the problem is reduced
to minimizing the response
to such noise while maintaining a constant response to ${\bf I_d}$. As a
result, the weights $\omega_j$ are the solution of the following system
of equations:
\begin{equation}
  \begin{array}{l}
    \vspace{0.2cm}
    1.\hspace{0.2cm}\boldsymbol{\omega}{\bf I_d^T}=1,\\
    2.\hspace{0.2cm}\partial(\boldsymbol{\omega}{\bf D}\boldsymbol{\omega}^T)/
       \partial\boldsymbol{\omega}=0,
    \end{array}
\end{equation}
where ${\bf D}=\langle{\bf S^T}{\bf S}\rangle-
\langle{\bf S^T}\rangle\langle{\bf S}\rangle$ is the data covariance
matrix and $\langle\rangle$ means averaging over the spatial pixels of the
sky map. The solution of such a system is:
\begin{equation}
  \boldsymbol{\omega}={\bf I_d^{}}{\bf D^{-1}}\cdot
  \left({\bf I_d^{}}{\bf D^{-1}}{\bf I_d^T}\right)^{\bf -1}.
\end{equation}

This completely blind approach can be biased due to nonzero
projections of foregrounds on the considered signal, 
${\bf F}{\bf I_d^T}\ne 0$. Taking into account that in our case the
foreground amplitudes can exceed the amplitude of the signal of interest
by several orders of magnitude, this method is not suitable.
\subsection{Moments approach (MILC)}

In order to avoid biasing in the process of signal ${\bf I_d}$ extraction,
the constrained ILC (or cILC) method was proposed \citep{2011MNRAS.418..467R, 2020MNRAS.494.5734R}, which nullifies
the response
to some modeled foregrounds with known spectra. Thus, a number of constraints
are added to Eq. (9) that ensure a zero response to certain foregrounds.
This modification of the blind ILC approach was
extended with the MILC method \citep{2005MNRAS.357..145S, 10.1093/mnras/stx1982, 10.1093/mnras/staa3292}, where foregrounds with badly
defined spectra in Eq. (3) were considered. The idea behind the MILC method is
quite simple and
effective. Since the spectral shapes of some signal components
depend on the parameters ${\bf P}$, they can be expanded in a Taylor series
up to some order $n$ in the vicinity of a certain reference point
corresponding to the average preestimated value ${\bf P_0}$:
\begin{equation}
  \begin{array}{c}
    \vspace{0.2cm}
    f_m(\nu_j,{\bf P})\approx f_m(\nu_j,{\bf P_0})+\\
    \sum\limits_{n_{{}_1},.,n_{{}_L}}
  \left[\frac{\partial^{n_{{}_1}+..+n_{{}_L}}}
    {\partial P_{{}_1}^{n_{{}_1}}\cdot\cdot P_{{}_L}^{n_{{}_L}}}
    f_m(\nu_j,{\bf P})\right]_{{}_{\bf P=P_0}}
  \Delta P_{{}_1}^{n_{{}_1}}\cdot\cdot\Delta P_{{}_L}^{n_{{}_L}},
  \end{array}
\end{equation}
where the summation is performed over all positive $n_{{}_1},..,n_{{}_L}$
satisfying the condition $0\le n_{{}_1}+..+n_{{}_L}\le n$. In order to remove
the influence of such foregrounds on data processing, it is sufficient to
require a zero response to all derivatives of the spectra with respect to
parameters up to the $n$th order. As a result, we get the following system
of equations: 

 \begin{equation}
    \begin{array}{l}
      \vspace{0.2cm}
     1.\hspace{0.2cm} \boldsymbol{\omega}{\bf I_d^T}=1,\\
       \vspace{0.2cm}      
 2.\hspace{0.2cm} \sum\limits_{j}\omega_j\frac{\partial^{n_{{}_1}+..+n_{{}_L}}}
    {\partial P_{{}_1}^{n_{{}_1}}\cdot\cdot P_{{}_L}^{n_{{}_L}}}
    f_m(\nu_j,{\bf P})|_{{}_{\bf P=P_0}}=0,\\
     \vspace{0.2cm}
  \hspace{0.4cm}m=1,..,M,\hspace{0.2cm}0\le n_{{}_1}+..+n_{{}_L}\le n,\\
  3.\hspace{0.2cm} \partial(\boldsymbol\omega{\bf C}\boldsymbol\omega^{{\bf T}})/
    \partial\boldsymbol{\omega}=0.
  \end{array}
 \end{equation}
 Therefore, this method ensures that the response to all foregrounds is
 zeroed to within
  \begin{equation}
    R({\bf F})=O\left[\sum\limits_j\omega_j\frac{\partial^{n+1}
        f_m(\nu_j,{\bf P})}
    {\partial{\bf P}^{n+1}}\right]_{{\bf P}={\bf P_0}}=R_{n+1}({\bf F}).
  \end{equation}

  However, the disadvantage of MILC is that with a large number of
  constraints in formula 2 of Eq. (12), the response to noise $R({\bf N})$ can
  become unacceptably large. Indeed, this formula is equivalent to the
  system of equations
  $\boldsymbol{\omega}\boldsymbol{\varphi_k^T}=0$,
  $k=1,..K$, where the vectors $\boldsymbol{\varphi_k^{}}$ correspond to the
  derivatives of the foregrounds with respect to the parameters ${\bf P}$
  and the capital $K$ denotes the total number of constraints (or derivatives).
  The system of vectors $\boldsymbol{\varphi_k^{}}$ can be orthogonalized
  in such a way that the new vectors
  $\boldsymbol{\tilde}{\boldsymbol{\varphi}}_{\bf k}^{}$
    are a linear combinations of
  the original vectors $\boldsymbol{\varphi_k^{}}$
  and represent an orthonormal system: 
  \begin{equation}
    \begin{array}{l}
      \vspace{0.2cm}
      \boldsymbol{\tilde}{\boldsymbol{\varphi}}_{\bf k}^{}=
      \boldsymbol{\tilde}{\boldsymbol{\varphi}}_{\bf k}^{}
      (\boldsymbol{\varphi}_{\bf 1}^{},..,\boldsymbol{\varphi}_{\bf K}^{}),\\
      
      \boldsymbol{\tilde}{\boldsymbol{\varphi}}_{\bf k}^{}
      \boldsymbol{\tilde}{\boldsymbol{\varphi}}_{\bf k'}^{\bf T}=
      \delta_k^{k'},\hspace{0.5cm}k,k'=1,...,K.           
    \end{array}
  \end{equation}
  Thus, the normalized signal ${\bf\tilde{I}_d}={\bf I_d^{}}\cdot
  ({\bf I_d^{}}{\bf I_d^T})_{}^{\bf -\frac{1}{2}}$
  can be written in the following form:
\begin{equation}
    \begin{array}{l}
      \vspace{0.2cm}
      {\bf\tilde{I}_d}=\sum\limits_{k=1}^K\gamma_{k}^{}
      \boldsymbol{\tilde}{\boldsymbol{\varphi}}_{\bf k}^{}+
      \boldsymbol{\Delta_d},\\
      
      \gamma_{k}^{}= \boldsymbol{\tilde}{\boldsymbol{\varphi}_k}
      \cdot{\bf\tilde{I}}_{{\bf d}}^{\bf T},     
    \end{array}
  \end{equation}
where $\boldsymbol{\Delta_d}$ is part of signal ${\bf\tilde{I}_d}$
orthogonal to all modeled foregrounds:
$\boldsymbol{\tilde}{\boldsymbol{\varphi}}_{\bf k}\boldsymbol
{\Delta_d^T}=0$, $k=1,..,K$. That is, the smaller the length
of the vector $\boldsymbol{\Delta}_{\bf d}$, the more difficult it is to separate
the vector ${\bf I_d}$ from the rest of the total signal.
We can define the value of
$\Gamma=\left(\boldsymbol{\Delta}_{\bf d}^{}
\boldsymbol{\Delta}_{\bf d}^{T}\right)_{}^{1/2}$
as a measure of the orthogonality of the normalized signal
${\bf\tilde{I}_d}$  to all foregrounds. It is easy to show, that
$\Gamma_{}^2=1-\sum\limits_{k=1}^K\gamma_k^2$. Finally, the solution
of the system of Eq. (12) for weights $\boldsymbol{\omega}$ is
\begin{equation}
  \boldsymbol{\omega}={\bf\Delta_d}{\bf C_{}^{-1}}\cdot
  \left({\bf\Delta_d}{\bf C^{-1}}{\bf I_d^T}\right)^{\bf -1},
\end{equation}
and the noise response can be estimated as follows:
\begin{equation}
  \begin{array}{l}
    \vspace{0.2cm}
  \langle R^2({\bf N})\rangle=\boldsymbol{\omega}{\bf C}
  \boldsymbol{\omega}^T\sim\sigma^2/\Gamma^2,\\
  \sigma^2=\frac{\langle{\bf N}{\bf N_{}^T}\rangle}{J}.
  \end{array}
\end{equation}
Here, $\sigma$ is equivalent to the mean sensitivity per frequency channel.
Thus, a small value of $\Gamma$ means a large response to random noise.
In case when $\Gamma\rightarrow 0$, the signal ${\bf I_d}$ becomes a linear
combination of foregrounds, and the separation of such a signal from
them with this method becomes impossible.

 \subsection{Least response method}

 Recently, a new approach to the component separation problem
 has been proposed \citep{2023PhRvD.107f3506N}, which can be called the LRM.
 Unlike MILC, this method implies the availability of information not only
 about the possible foreground spectral shapes variations, but also about their
 maximum possible amplitudes. It is important to note that such information
 about main known foregrounds is available to us. In addition, overestimation of the
 upper limit of the foreground amplitudes is not critical for our approach
 (unlike underestimation).
 This helps to avoid imposing strict
 conditions on the weight vector $\boldsymbol{\omega}$ and thereby
 significantly reduce
 the noise response. The only fairly mild assumption about the foregrounds
 described in Eq. (3) is that the amplitudes $a_m$ inside the
 parameter domain $\Omega$
 should be less than certain (preestimated) values $A_m$:
\begin{equation}
  \mid a_m({\bf P})\mid \le A_m \hspace{0.3cm}for\hspace{0.3cm}{\bf P}
  \in \Omega,
\end{equation}
and $a_m({\bf P})=0$ otherwise. 

It is easy to show, that the mean square of the response to foreground
has an upper limit:
\begin{equation}
    \langle R^2({\bf F})\rangle\le
    \langle\sum\limits_{m=1}^MM\cdot a_m^2({\bf P})\left[\sum\limits_{j=1}^J
    f_m(\nu_j,{\bf P})
      \cdot\omega_j\right]^2\rangle,
\end{equation}
and according to Eq. (18) the following inequality is always true:
\begin{equation}
  \begin{array}{l}
     \vspace{0.3cm}
    \langle R^2({\bf F})\rangle\le
    \boldsymbol\omega\boldsymbol\Phi\boldsymbol\omega^{{\bf T}},
    \hspace{0.3cm}
    \boldsymbol\Phi=M\left[\sum\limits_{m=1}^MA_m^2q_{ij}^m\right],\\
    q_{ij}^m=\frac{1}{V_\Omega}\int\limits_{\Omega}
    f_m(\nu_i,{\bf P})f_m(\nu_j,{\bf P})d{\bf P}
 \end{array}   
\end{equation}
where integrals $q_{ij}^m$ can be precalculated for all types of foreground
($m=1,..,M$) numerically or in some particular cases analytically depending
on the configuration of $\Omega$. It is worth noting that the prefactor $M$
for the matrix ${\bf\Phi}$ in Eqs. (19) and (20) is used to ensure that these
inequalities are certainly correct. In general, the prefactor for this matrix can vary from 1 (when there are no correlations between foregrounds) to M (for 100\% correlation between all foregrounds). Thus, using the value M as a prefactor provides a complete guarantee of taking into account all possible correlations. In reality, foregrounds of different
physical origins are weakly correlated with each other. In this case, the
coefficient M can be replaced by one. In our estimates we use uncorrelated
foregrounds model and, therefore,
$\boldsymbol\Phi=\left[\sum\limits_{m=1}^MA_m^2q_{ij}^m\right]$. For a more
detailed analysis, possible correlations between individual foregrounds can
be taken into account in the calculation of the ${\bf\Phi}$ matrix.

Since $\langle R^2({\bf N})\rangle=
\boldsymbol\omega{\bf C}\boldsymbol\omega^{{\bf T}}=\sum\limits_{i,j}C_{ij}\omega_i\omega_j$ and the foregrounds are not correlated with noise we have

\begin{equation}
 \langle\left(R({\bf F})+R({\bf N})\right)^2\rangle\le
 \boldsymbol\omega[\boldsymbol\Phi+{\bf C}]\boldsymbol\omega^{{\bf T}}.
\end{equation}
Therefore the minimization of the response to the foreground and to the noise
is achieved with weights $\omega_j$ corresponding to the minimum of the
quadratic form
$\boldsymbol\omega[\boldsymbol\Phi+{\bf C}]\boldsymbol\omega^{{\bf T}}$
under condition in Eq. (7): 

\begin{equation}
    \begin{array}{l}
      \vspace{0.2cm}
     1.\hspace{0.2cm} \boldsymbol{\omega}{\bf I_d^T}=1,\\
       \vspace{0.2cm}      
     2.\hspace{0.2cm} \partial(\boldsymbol\omega\left[{\bf\boldsymbol\Phi+C}
       \right]
       \boldsymbol\omega^{{\bf T}})/\partial\boldsymbol\omega=0.
  \end{array}
 \end{equation}
The solution of the system Eq. (22) is
\begin{equation}
    \boldsymbol{\omega}={\bf I_d}\boldsymbol{ \left[ } {\bf\Phi}+{\bf C}
    \boldsymbol{\right]}^{{}^{-1}}\cdot
  \left({\bf I_d}\boldsymbol{\left[}{\bf\Phi}+
    {\bf C}\boldsymbol{\right]}^{{}^{-1}}{\bf I_d}^{{}^T}\right)^{-1}.
\end{equation}

Thus, LRM, unlike MILC, does not require complete orthogonality of the signal of
interest to all foregrounds [compare Eqs. (12) and (22)].
In the next section we will demonstrate the advantages of this approach and
show that it allows to detect a signal in the observational data at a much
lower sensitivity.
 
\section{Methods comparison and prospects for measuring spectral distortions}
\subsection{Modeling spectral distortions and foregrounds}
Our numerical calculations were performed for a simulated observed frequency spectrum that included the signal
of interest, all possible foregrounds with varying parameters and photon noise. We did not use any information
about the spatial distribution of foreground sources in the
sky. In order to carry out a numerical experiment on the application of the two
methods (MILC and LRM) described in the previous section, we used the
following models of signal components.

{\it Signals of CMB origin} are $\mu$ distortion $I_\mu$, $y$ distortion
(SZ effect) $I_{y_0}$, first and second relativistic corrections to thermal
SZ effect $I_{y_1}$, $I_{y_2}$ and CMB anisotropy $I_{{}_{CMBA}}$.
(It is assumed that the CMB monopole can be easily removed from the
observational data).
All these spectra have well-defined shapes that do not depend on any
parameters and look as follows: 
\begin{equation}
   \begin{array}{l}
    \vspace{0.2cm}
    I_\mu(\nu)=I_0\frac{x^4e^x}{(e^x-1)^2}\left(\frac{1}{b}-\frac{1}{x}
    \right)\mu,\\
    \vspace{0.2cm}
    I_{y_0}(\nu)=I_0\frac{x^4e^x}{(e^x-1)^2}
    \left(x \coth(\frac{x}{2})-4\right)y,\\
    \vspace{0.2cm}
    I_{y_{1}}(\nu)=I_0\frac{x^4e^x}{(e^x-1)^2}Y_1(x)
    \hspace{0.1cm}\theta_e\hspace{0.05cm}y,\\
    \vspace{0.2cm}
    I_{y_{2}}(\nu)=I_0\frac{x^4e^x}{(e^x-1)^2}Y_2(x)
    \hspace{0.1cm}\theta_e^2\hspace{0.05cm}y,\\
    I_{{}_{CMBA}}(\nu)=\frac{2(kT_0)^3}{(hc)^2}\frac{x^4}{(e^x-1)^2}
    \hspace{0.1cm}\frac{\Delta T}{T_0}
    \end{array}
\end{equation}
where $x=h\nu/kT_{{}_{CMB}}$ and the CMB temperature is $T_{{}_{CMB}}=2.72548$ $K$, \citep{1990ApJ...354L..37M,Fixsen_2009}
and $\Delta T/T_{{}_{CMB}}<10^{-4}$.
The same estimated values for constants $b$, $I_0$, $\mu$ and $y$ as in
\citep{2017MNRAS.471.1126A} are used: $I_0=270$ MJy/sr, $b = 2.1923$, $\mu=2\times 10^{-8}$,
$y=1.77\times 10^{-6}$ and $\theta_e=kT_{{}_{SZ}}/m_ec^2\sim 2.44\times 10^{-3}$.
The functions $Y_1(x)$ and $Y_2(x)$ for the first and second corrections have
a rather cumbersome form and analytical formulas for them can be found in
\citep{2000MNRAS.312..159C}.

{\it Dust and CIB foregrounds} are considered together and modeled as a
modified blackbody radiation with two floating parameters: temperature $T$
and spectral index $\beta$,
\begin{equation}
  \begin{array}{l}
    \vspace{0.2cm}
  I_{{}_{Dust,CIB}}(\nu,T,\beta)=\tau_{{}_{DC}}(\nu/\nu_{{}_{DC}})^\beta B(\nu,T),\\
  B(\nu,T)=\frac{2(kT)^3}{(hc)^2}\frac{x^3}{e^x-1},
  \end{array}
\end{equation}
where $\nu_{{}_{DC}}=353$ GHz.
The boundaries of the parameters $(T,\beta)$ domain were determined in \citep{2023PhRvD.107f3506N}
using Planck data \citep{2014A&A...571A..11P,2016A&A...596A.109P}.
The probability distribution function for these parameters was calculated
for 10 degrees circular sky part centered at
$l=13.731^o$, $b=-73.946^o$, see Fig. 1(a).
Two isocontour solid lines limit the region $\Omega(T,\beta)$ of possible
parameters variations for both dust and CIB. The probability to find
parameters outside these two spots is less than 0.0002.
The maximum allowable value of emissivity $\tau_{{}_{DC}}$ for the data
we used does not exceed $10^{-6}$.

{\it Synchrotron radiation} is modeled according to \citep{1986rpa..book.....R} and its spectrum
has a power-law form with the single free parameter $\beta_s$: 
\begin{equation}
  I_{{}_{sync}}(\nu,\beta_s)=A_s(\nu/\nu_s)^{-\beta_{s}},\\
\end{equation}
where $\nu_s=30$ GHz and $A_s<1000$ Jy/sr. In accordance with the results
of \citep{delahoz23}, $\beta_s$ can vary from 0.9 to 1.4.

{\it Free-free emission} is given by the following formula in \cite{2016A&A...594A..10P}:
\begin{equation}
  I_{{}_{ff}}(\nu)=A_{ff}\left(1+\ln\left[1+\left(\frac{\nu_{ff}}{\nu}\right)
    ^{\sqrt{3}/\pi}\right]\right),
\end{equation}
where, $A_{ff}<500$ Jy/sr,
$\nu_{ff}=255.33\left(\frac{T_e}{1000 K}\right)^{3/2}$ GHz.
According to \citep{2016A&A...594A..10P}, the parameter $T_e=7000K\pm 3K$
does not vary across the sky strongly enough to noticeably make any difference
in the shape of the spectrum. Therefore, in our calculations we consider
free-free spectrum as well-defined one without any parameter variations.

{\it The instrument optics emission} is considered in this paper as a graybody
radiation with varying temperature $T_{opt}$:
\begin{equation}
  I_{opt}=\tau_{opt}B(\nu,T_{opt})
\end{equation}  
Temperature variations can depend on
the stability of the cooling system, design features and quality of the optical system.
In our numerical calculations we use a possible range of temperature
$T_{opt}$ variations  as $T_{min}\le T_{opt}\le T_{max}$ and $T_{max}-T_{min}\le
2 K$.
The emissivity $\tau_{opt}$ depends on the quality of the polishing of the
reflecting surfaces of the optical system. We use three different upper
limits for emissivity: $\tau_{opt}\le 0.001$,
$\tau_{opt}\le 0.01$ and $\tau_{opt}\le 0.05$.

{\it The photon noise} comes from the CMB radiation, a number of main foregrounds and emission from the instrument. All these sources of fluctuations can be characterized by the noise equivalent power ($NEP$) which is measured in units of W/Hz$^{1/2}$. For the FTS, intensity of the noise (i.e. the 1$\sigma$ sensitivity) is given by the following equation \citep{2012A&A...538A..86D}:

\begin{equation}
\sigma=0.61 \frac{ NEP}{\Delta\nu \sqrt{t} G},
\label{eq:sigma}
\end{equation}
 where $t$ is the integration time and
 $G$ is the throughput of the system. The value $G$ is the solid angle of the
 entrance pupil seen from the detector multiplied by the area of the
 instrument feedhorn. For the diffraction-limited instrument it can be
 estimated as
\begin{equation}
G_{dif} = \left(0.61\pi c/\nu_{min}^{}\right)^2.
\label{eq:throughput}
\end{equation}
However, the instruments we analyze are not necessarily diffraction-limited, so generally it is possible to have $G\gg G_{dif}$.

The $NEP$ is computed as
\begin{equation}
NEP_{}^{{}^2}=NEP_{det}^{{}^2}+NEP_{{}_F}^{{}^2},
\end{equation}
where $NEP_{det}^{}$ is the intrinsic $NEP$ of the detector while $NEP_{{}_F}^{}$
is created by $M$ various foregrounds [see Eq. (1)]. 
We assume that $NEP_{det}^{}\le$ $10^{-19} $ W$\cdot$Hz$^{-1/2} \ll NEP_{{}_F}$ in our
estimates. Therefore:
\begin{equation}
  \begin{array}{l}
    \vspace{0.2cm}
    \hspace{-0.3cm}NEP_{}^{{}^2}\approx NEP_{{}_F}^{{}^2}=
    \sum\limits_{m=1}^MNEP_m^{{}^2},\\
    
    \hspace{-0.3cm}NEP_{m}^{{}^2}=
    \int\limits_{\nu_{min}}^{\nu_{max}}\frac{4Gh^2}{c^2}
    \nu^{4}n_m(\nu) \left[ 1+ n_m(\nu)\right]d\nu,
  \end{array}
 \label{eq:NEP}
\end{equation}
where $n_m(\nu)=\frac{c^2}{2h\nu^3}I_m(\nu)$ is the photon concentration in phase space for $m$th foreground component. Here we assume ideal optical
and main beam efficiency.

If the noise is determined by a weak constant foreground $I(\nu)=const$ with $n_m \ll 1$,
from (\ref{eq:NEP}) and (\ref{eq:sigma}) one can obtain:
\begin{equation*}
\sigma \propto \sqrt{\frac{I}{t}}\frac{\nu_{max}-\nu_{min}}{\Delta \nu},
\end{equation*}
which coincides with the often used estimate of FTS sensitivity, see, e.g. \citep{2013ExA....35..527M}.

As one can see from Eqs. (29)-(32), the noise amplitude depends on the intensity
of radiation coming from the sky and from the all optical system, on the FTS frequency range and
on the spectral resolution. The main sources of the noise are
CMB monopole, dust radiation, CIB and the radiation emitted by the
instrument optics.
In this section we use a white
noise model. Therefore the noise covariance matrix has a diagonal form
${\bf C}=\sigma^2{\bf E}$, where $\sigma$ is the sensitivity. This kind
of noise distribution corresponds to a single
frequency band of the FTS instrument.

We consider the single band receiver
for 384 channels (7.5 GHz each) from 15 GHz to 2895 GHz.
The throughput is computed as for a diffraction-limited system, i.e.
from Eq. (30).

\subsection{Numerical results}

\subsubsection{Dust and CIB foregrounds}

\begin{figure*}[!htbp]
  \includegraphics[width=0.329\textwidth]{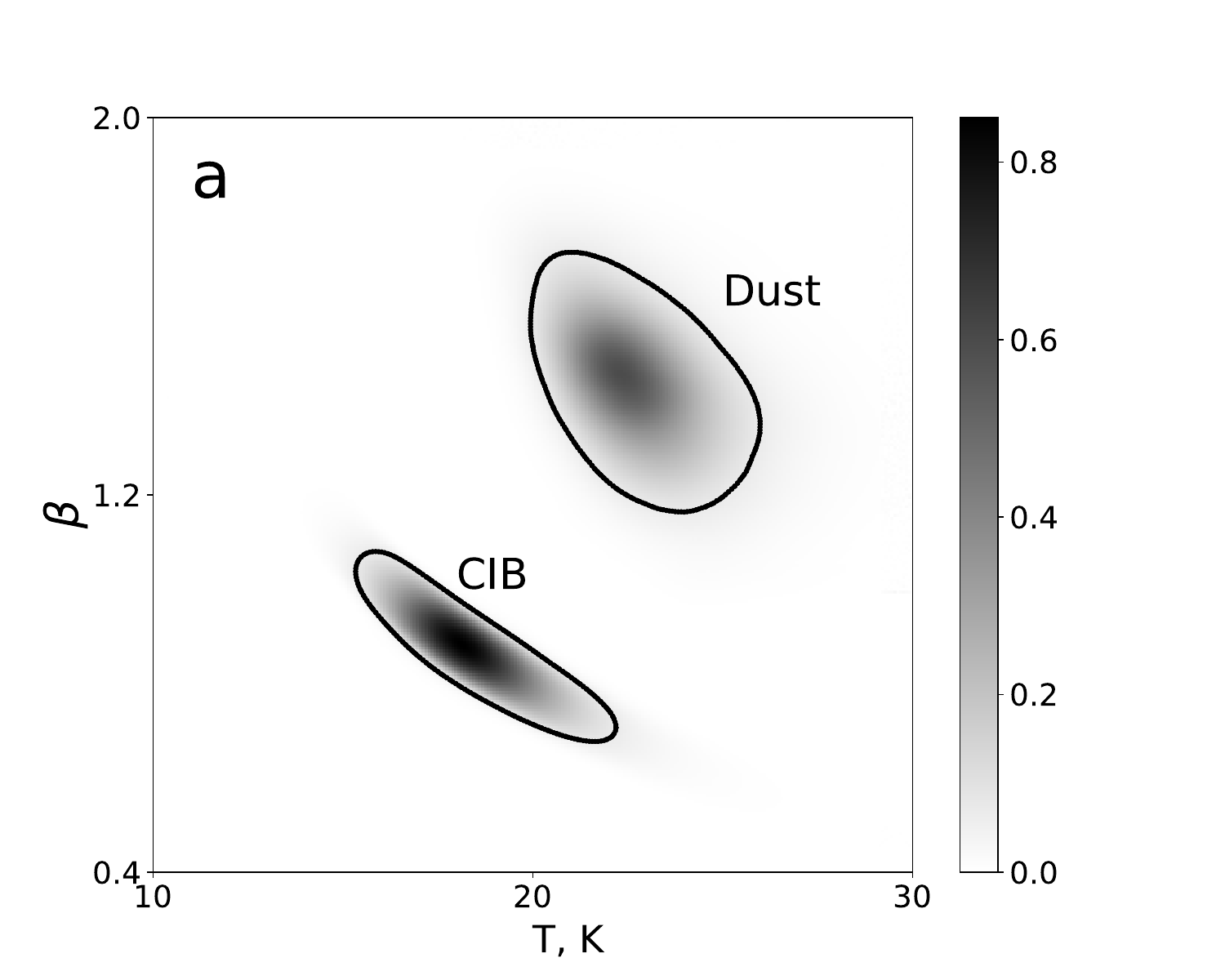}
  \includegraphics[width=0.329\textwidth]{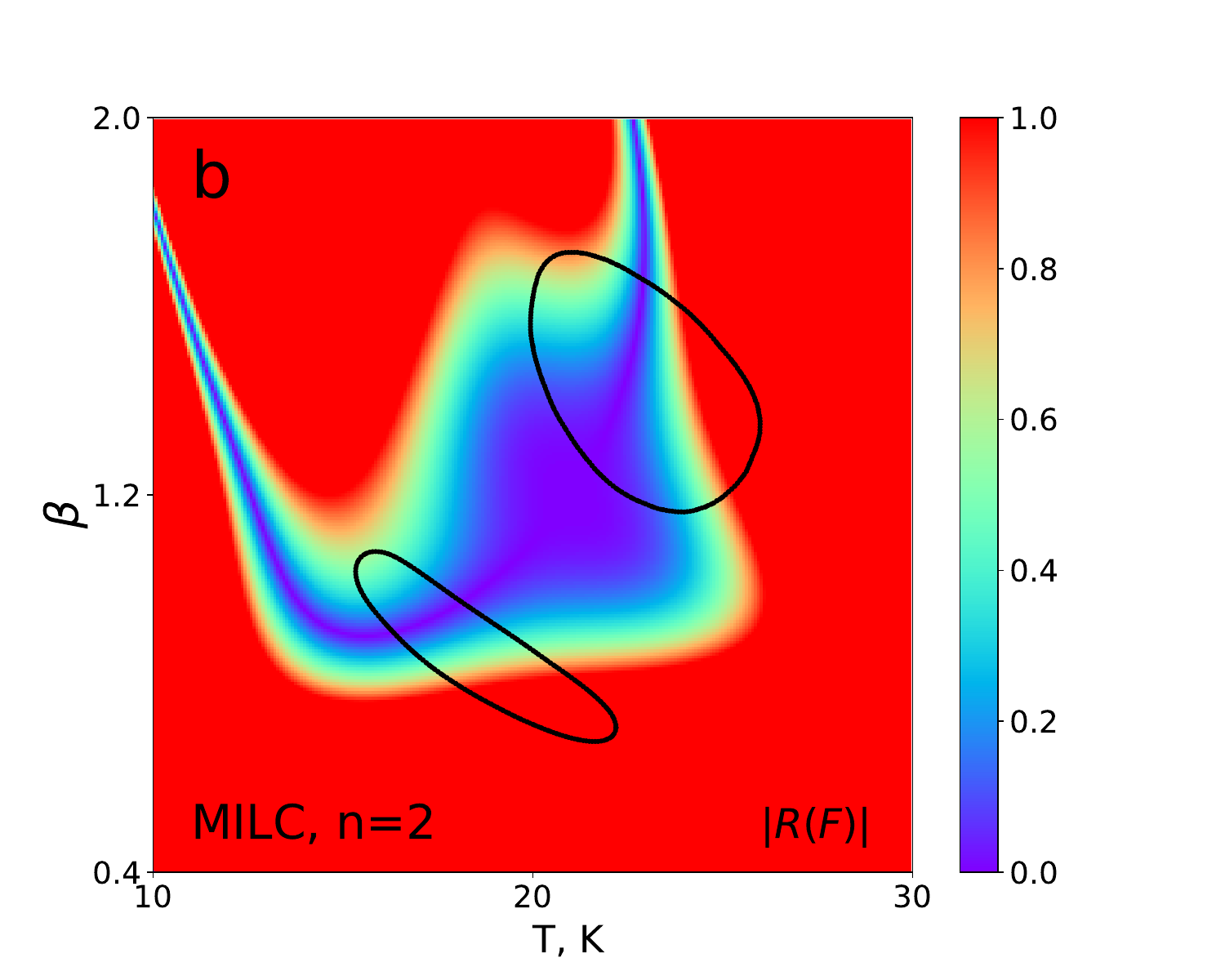}
  \includegraphics[width=0.329\textwidth]{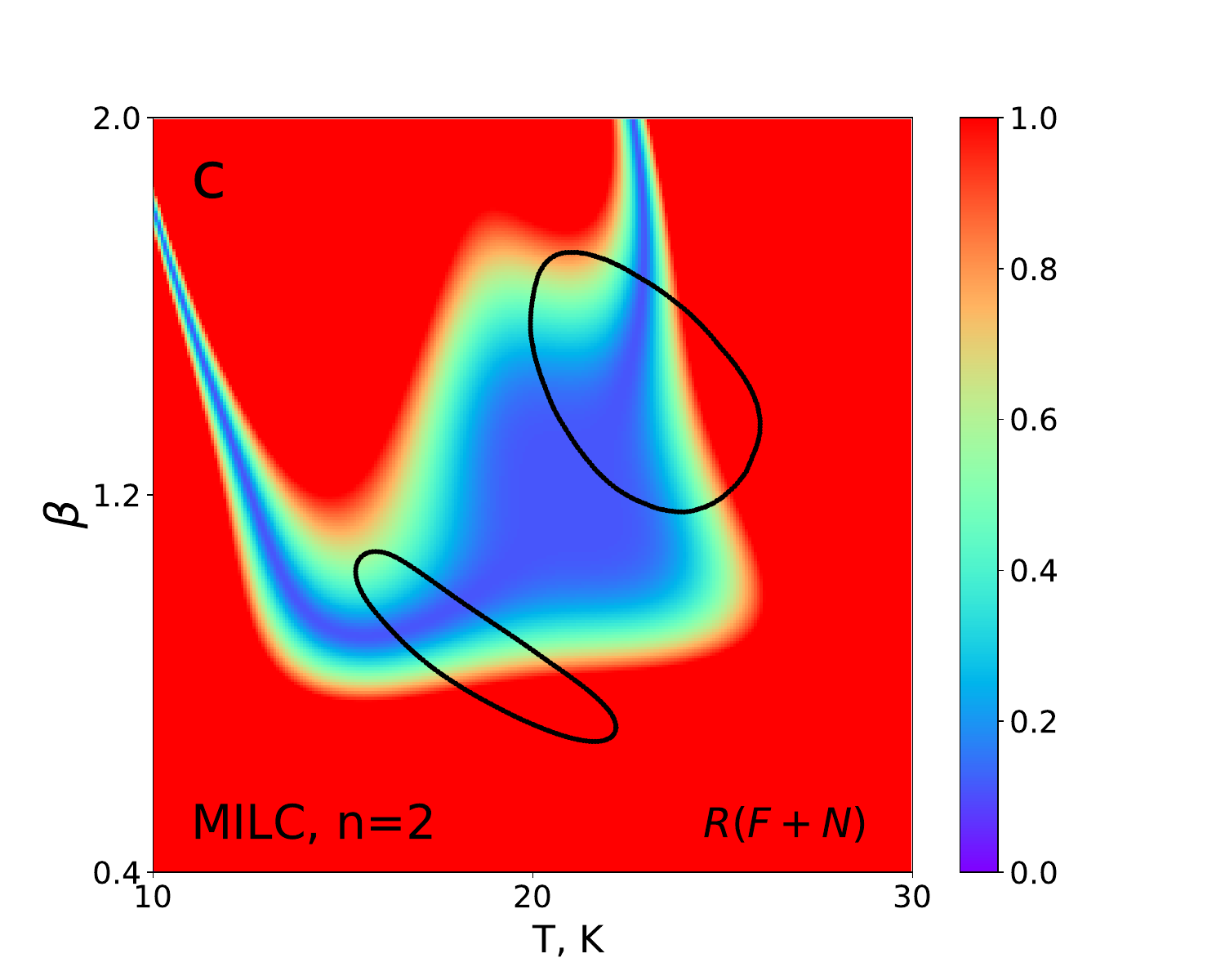}
  \includegraphics[width=0.329\textwidth]{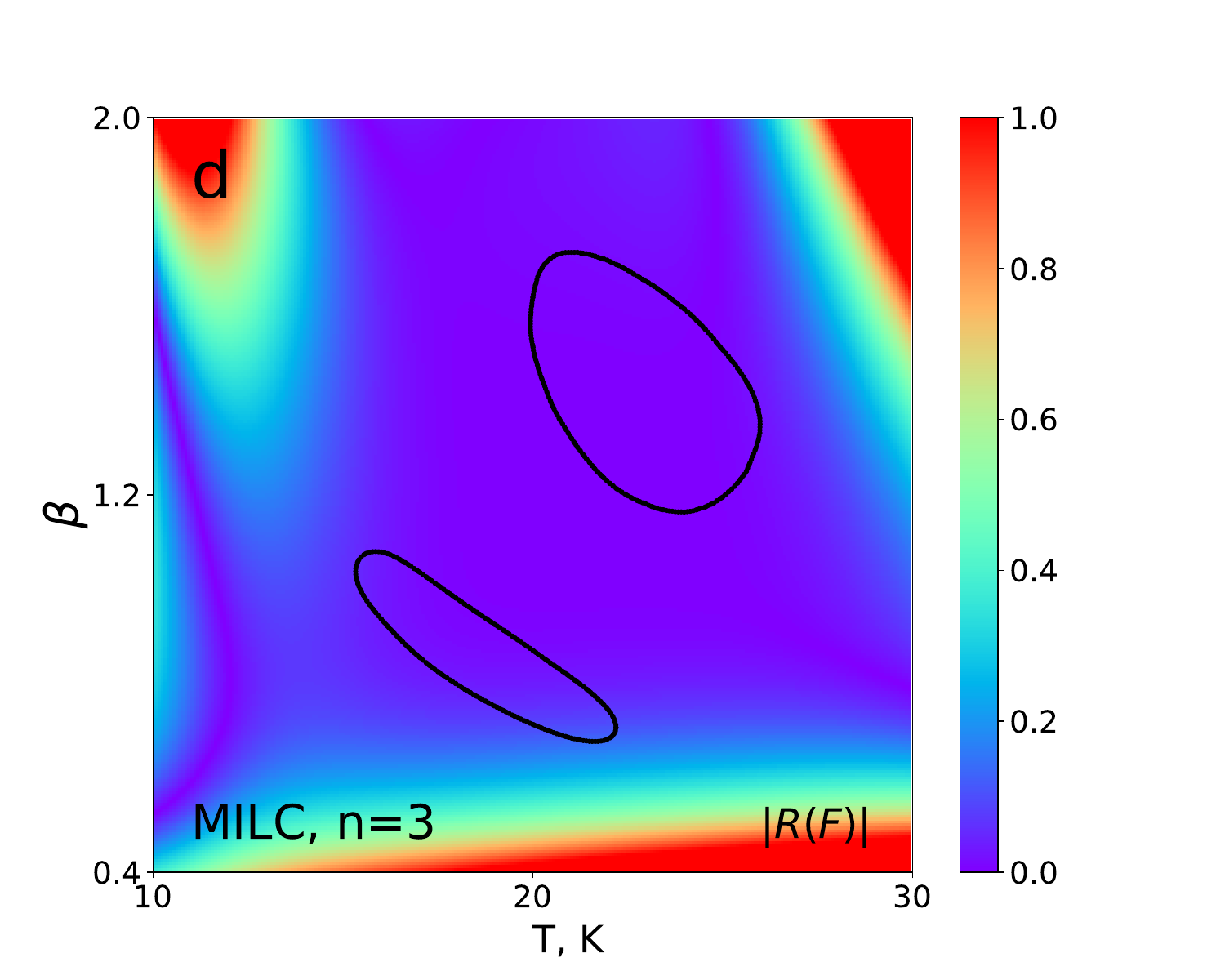}
  \includegraphics[width=0.329\textwidth]{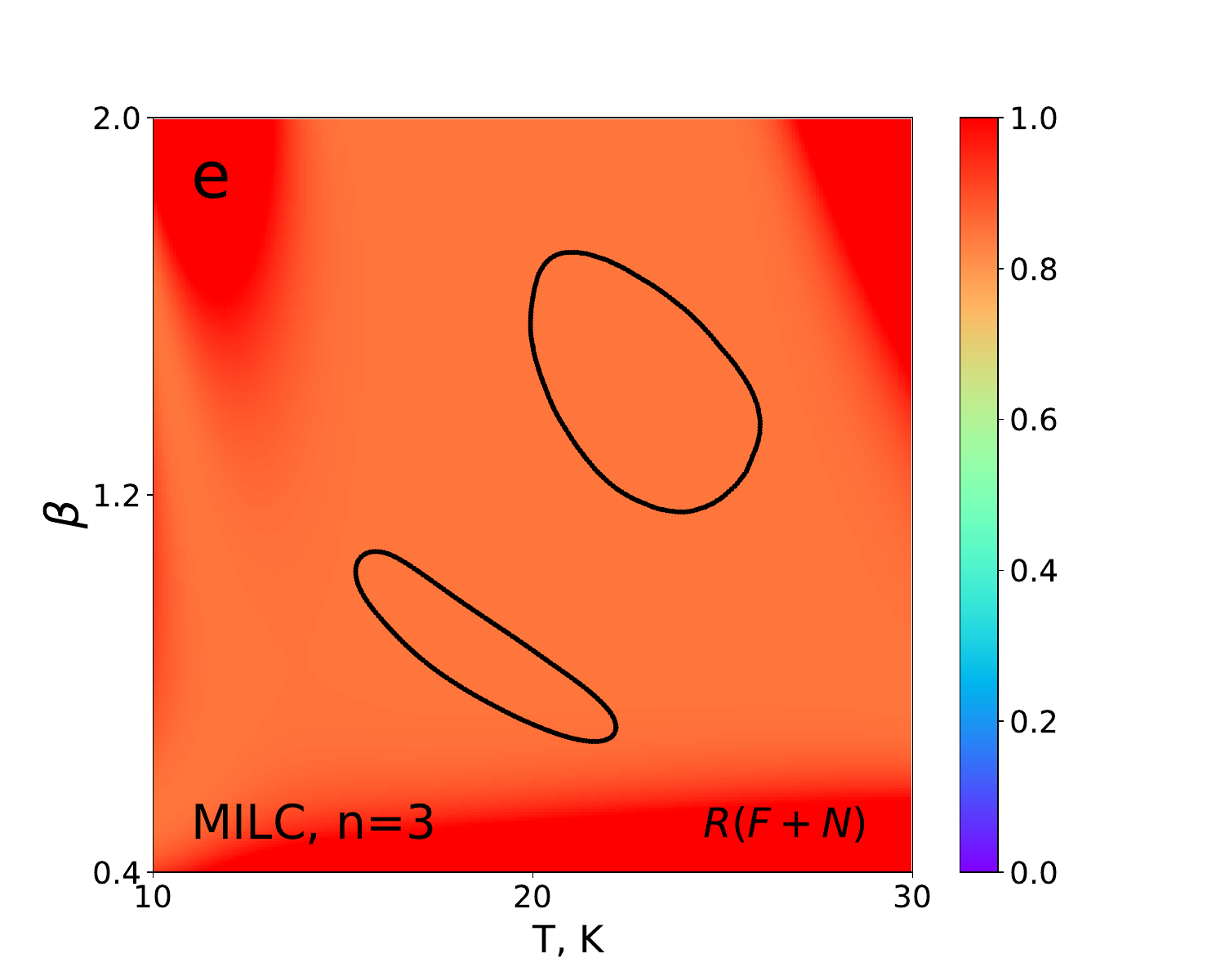}
  \includegraphics[width=0.329\textwidth]{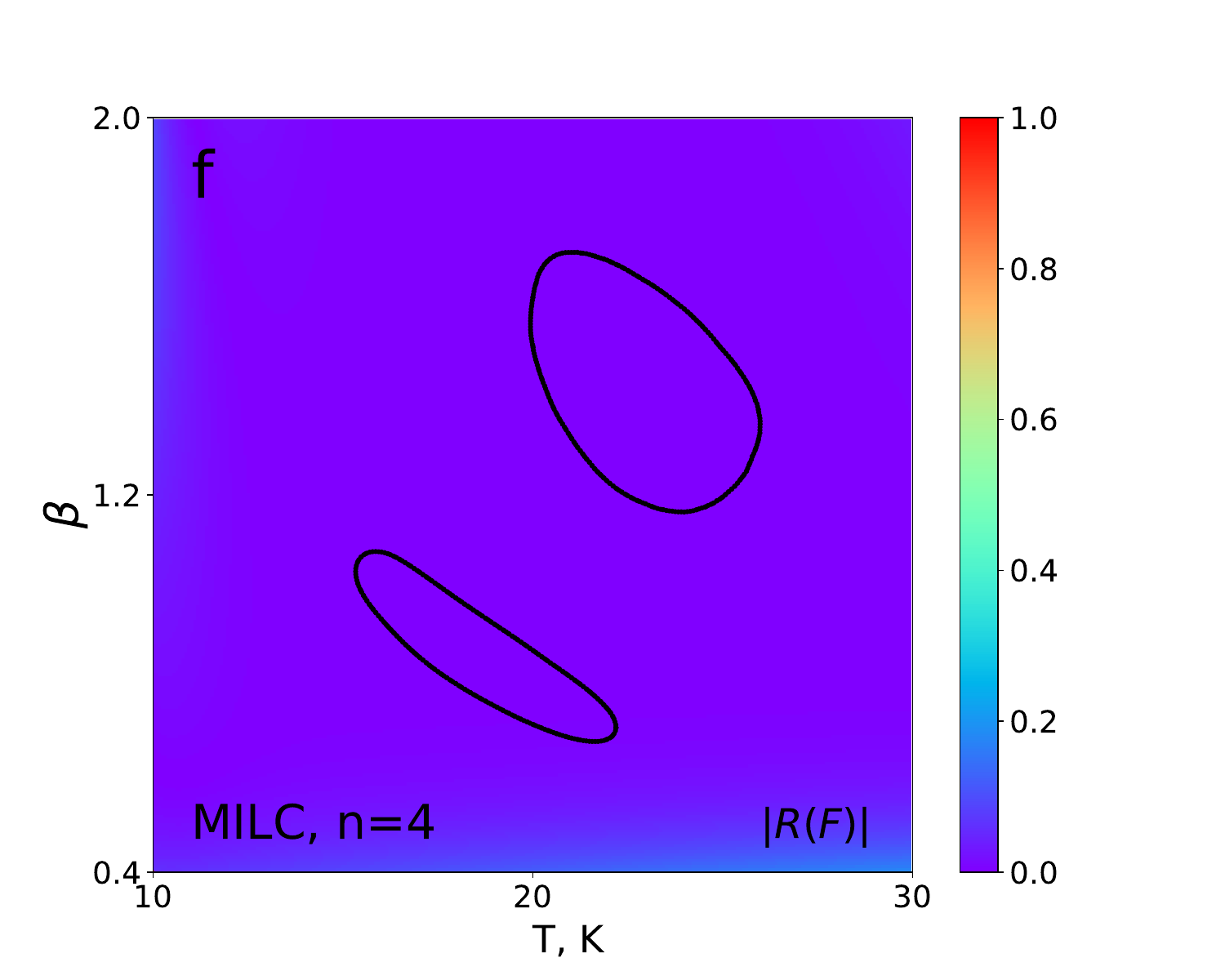}
  \includegraphics[width=0.329\textwidth]{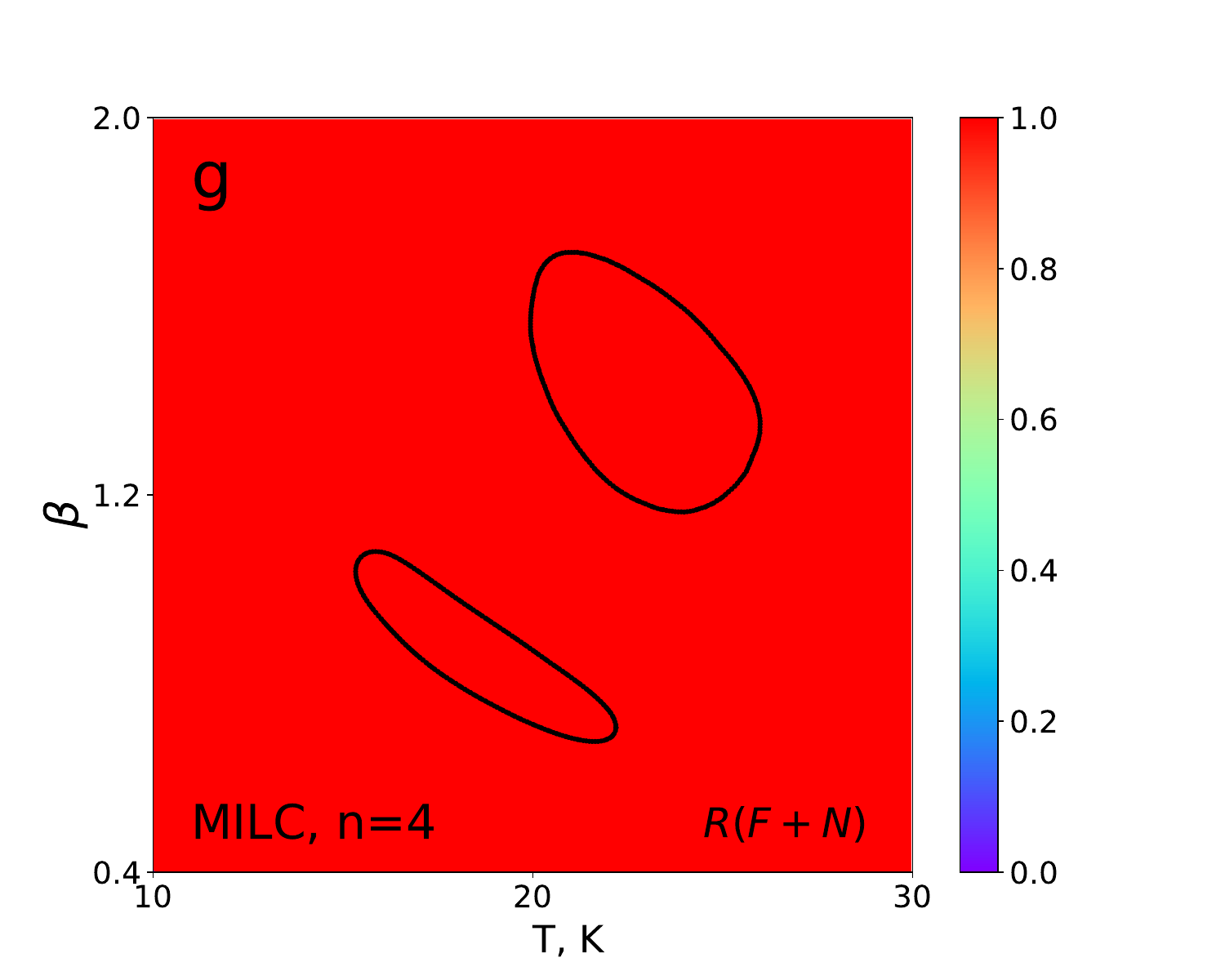}
  \includegraphics[width=0.329\textwidth]{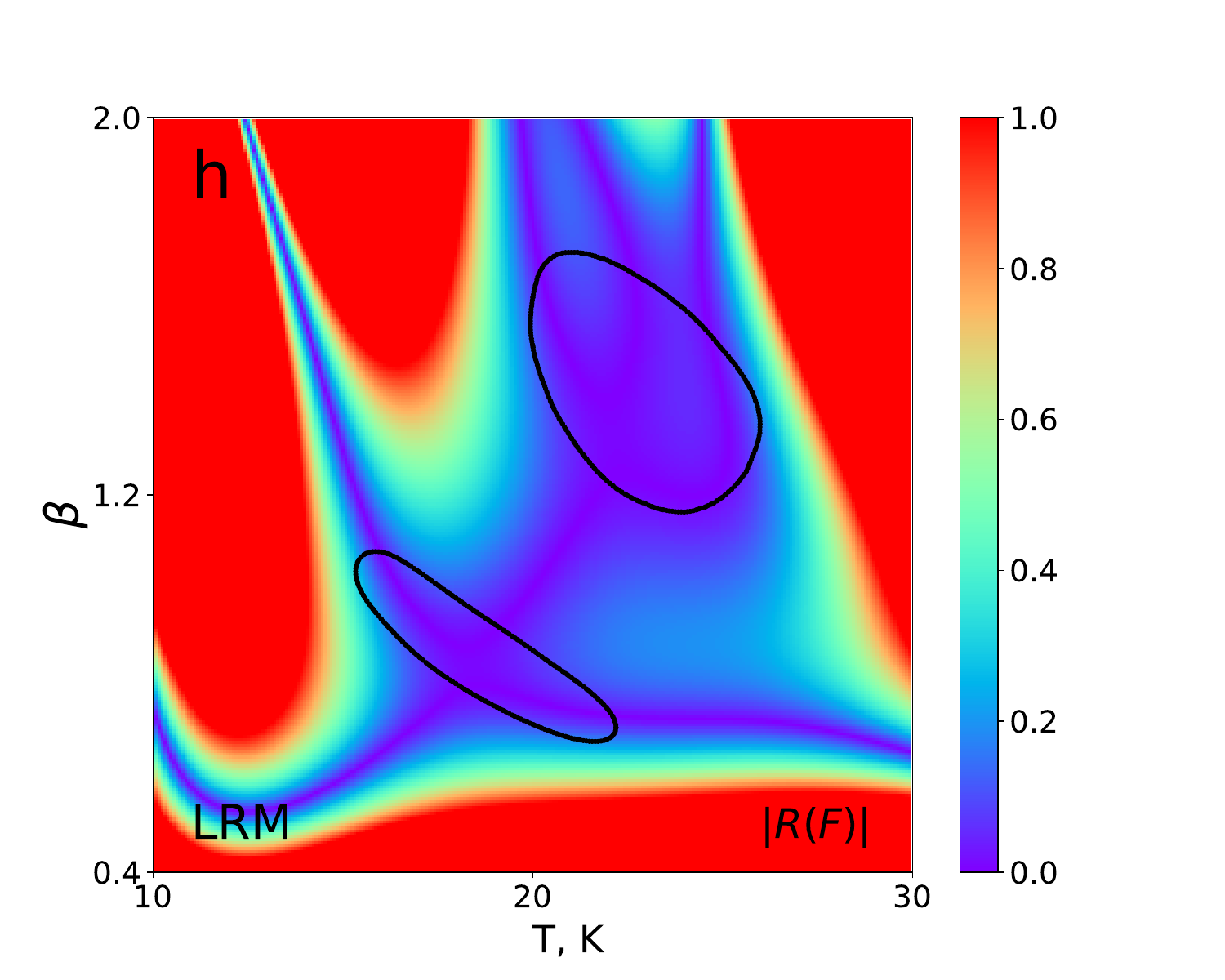}
  \includegraphics[width=0.329\textwidth]{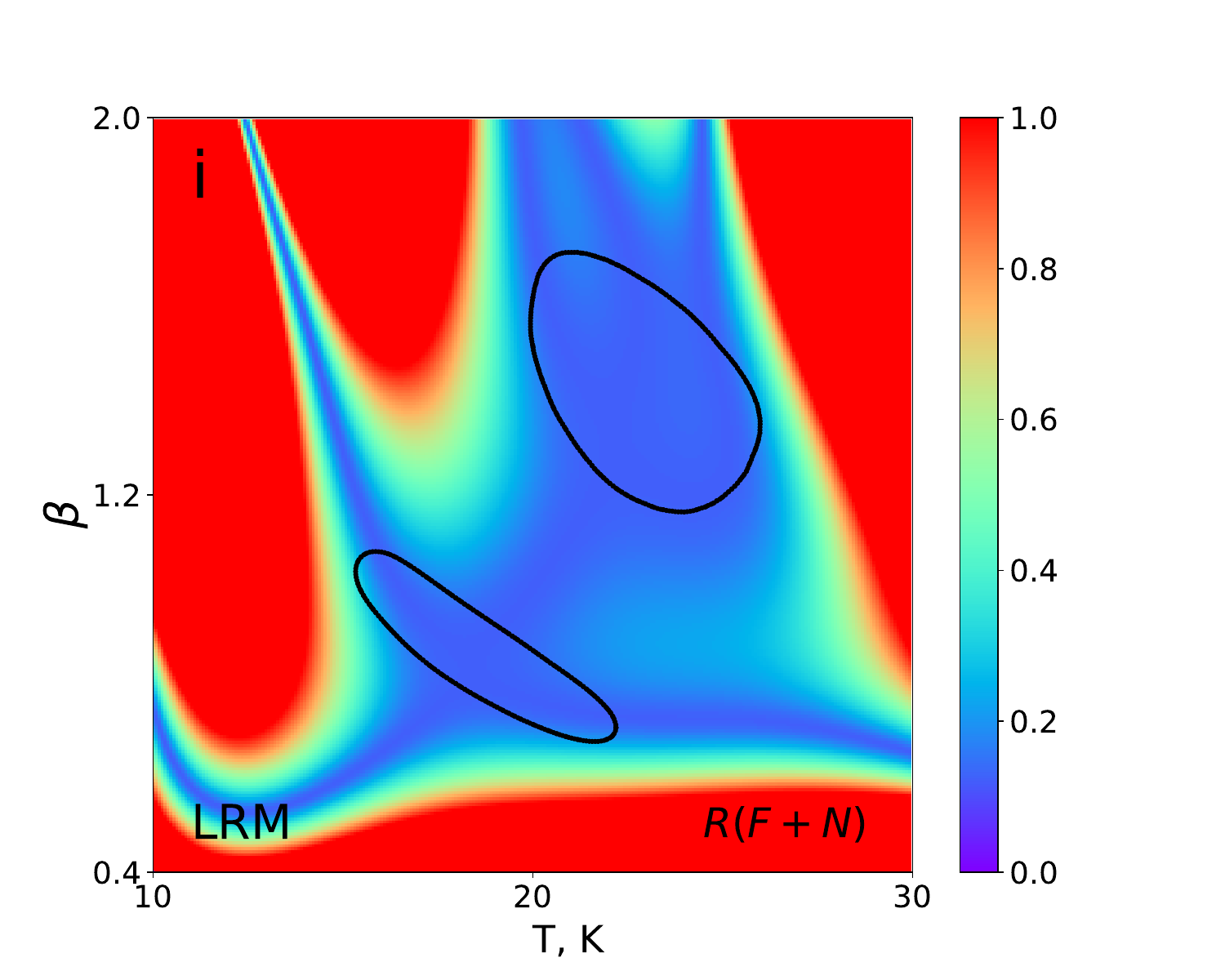}
  \caption {MILC and LRM methods for separating $\mu$
    distortions when only dust and CIB are taken as foregrounds. {\it Panel (a)}
    shows the probability distribution function for $T$ and $\beta$
    parameters. Isocontour lines limit the $\Omega$ region of parameter
    variations. {\it Panel (b)} shows MILC response to
    foreground $\mid R({\bf F})\mid$ if n=2. Dark red
    indicates the region where the response to the foreground exceeds the
    response to the $\mu$ signal: $\mid R({\bf F})\mid\ge 1$. 
    {\it Panel (c)} shows the total MILC noise+signal response
    $R({\bf F}+{\bf N})$ if n=2. {\it Panel (d):}
    $\mid R({\bf F})\mid$ for MILC, n=3. {\it Panel (e):}
    $R({\bf F}+{\bf N})$ for MILC, n=3. {\it Panel (f):}
    $\mid R({\bf F})\mid$ for MILC, n=4. {\it Panel (g):}
    $R({\bf F}+{\bf N})$ for MILC, n=4. {\it Panel (h):}
    $\mid R({\bf F})\mid$ for LRM. {\it Panel (i):}
    $R({\bf F}+{\bf N})$ for LRM.
  }
\end{figure*}
We start our analysis by comparing the efficiency of the MILC and LRM methods
for $I_\mu(\nu)$ distortion detection with a simple example when only dust and
infrared radiation are taken into account as foregrounds.
Noise level (or sensitivity) in this example is
$\sigma=1$ Jy/sr per single channel.
The MILC method was applied for three different versions of the Taylor
series expansion of the modified black body in terms of parameters
$T$ and $\beta$ [see Eq. (25)] to the second, third and fourth order respectively
in the vicinity of the reference value ${\bf P_0}$ of the vector
${\bf P}(T,\beta)$: ${\bf P_0}=(T_0,\beta_0)$, $T_0=21.2$ K $\beta_0=1.11$ in
Eq. (12). So the number of constraints
zeroing derivatives over two variables up to the $n$th order
are as follows: 6 constraints for n=2, 10 constraints
for n=3 and 15 constraints for n=4. The results are shown in Fig. 1.

For $n=2$, the response to foreground
$\mid R({\bf F})\mid$ and the total response to foreground and noise
$R({\bf F}+{\bf N})$ are shown in Fig. 1(b) and in Fig. 1(c) respectively in
comparison with the response to the $\mu$ signal
$R({\bf I_\mu})=1$. Obviously, the constraints imposed in Eq. (12)
for $n=2$ are not enough to clean the area $\Omega$ from the influence of
the foreground, that is, to ensure the condition
$\mid R({\bf F})\mid<<R({\bf I_\mu})$ inside two spots limited by isocontour
lines.

For $n=3$, the foreground response is quite low compared to the signal response,
although the noise response becomes relatively large and as a result
$R({\bf F}+{\bf N})\approx 0.8R({\bf I_\mu})$, see Fig. 1(d) and Fig. 1(e).

In the case when $n=4$ [Figs. 1(f) and 1(g)], the region $\Omega$ is excessively
clean from
the influence of the foreground, but at the same time 15 constraints
in Eq. (12) lead to an
extremely high response to random noise for the given sensitivity,
$\mid R({\bf N})\mid>>R({\bf I_\mu})$.

Thus, the choice of $n=3$ is the optimal number of moments in the decomposition
of dust and CIB spectra when using the MILC method. Figure 2 shows the
dependence of responses to foreground and noise on the number of constraints
imposed. As this number increases, the response to the foreground
gets smaller and, at the same time, the response to random noise rises
due to a decrease in the index $\Gamma$ which is a measure of orthogonality between foreground and $\mu$ signal [see Eq. (17)]. So the total response
to the foreground+noise reaches its minimum at a certain number
of constraints.

\begin{figure*}[!htbp]
  \includegraphics[width=0.49\textwidth]{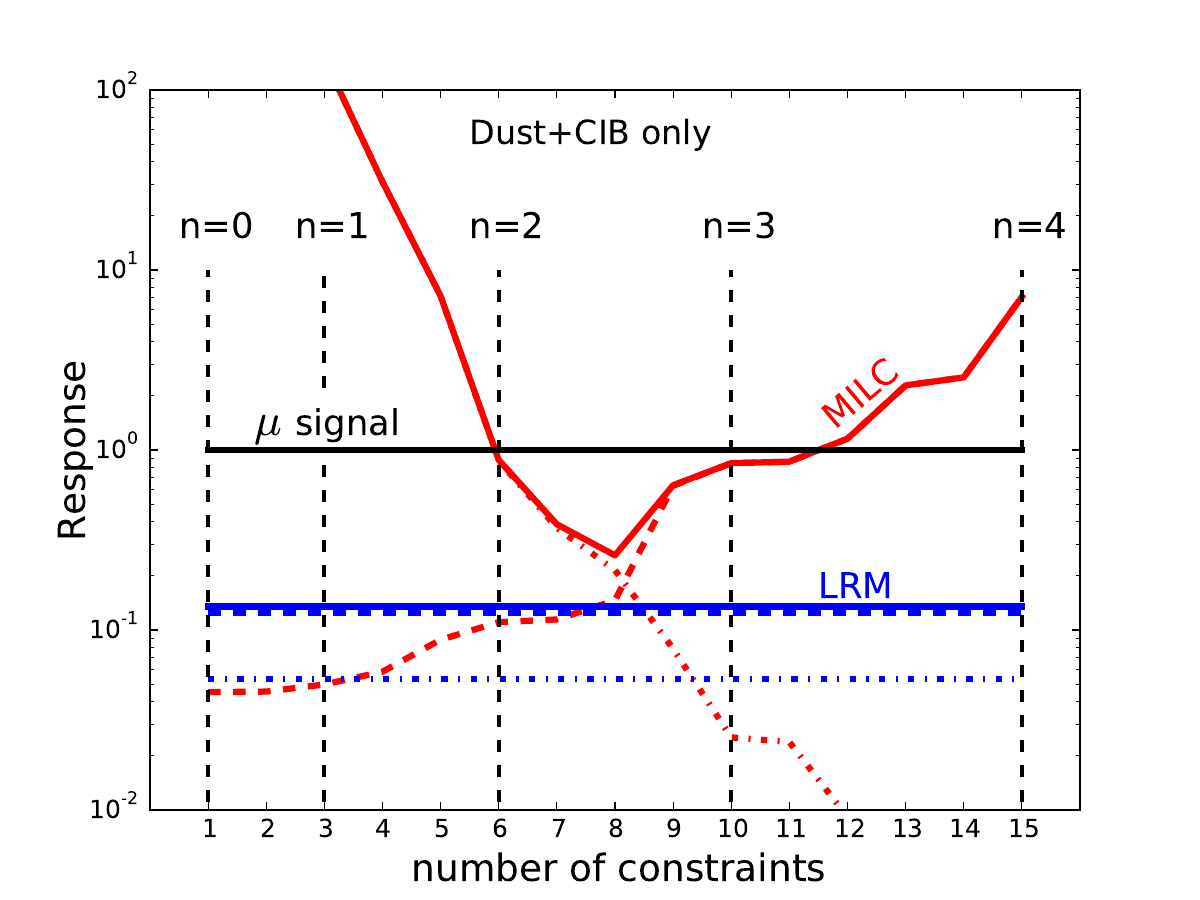}
  \includegraphics[width=0.49\textwidth]{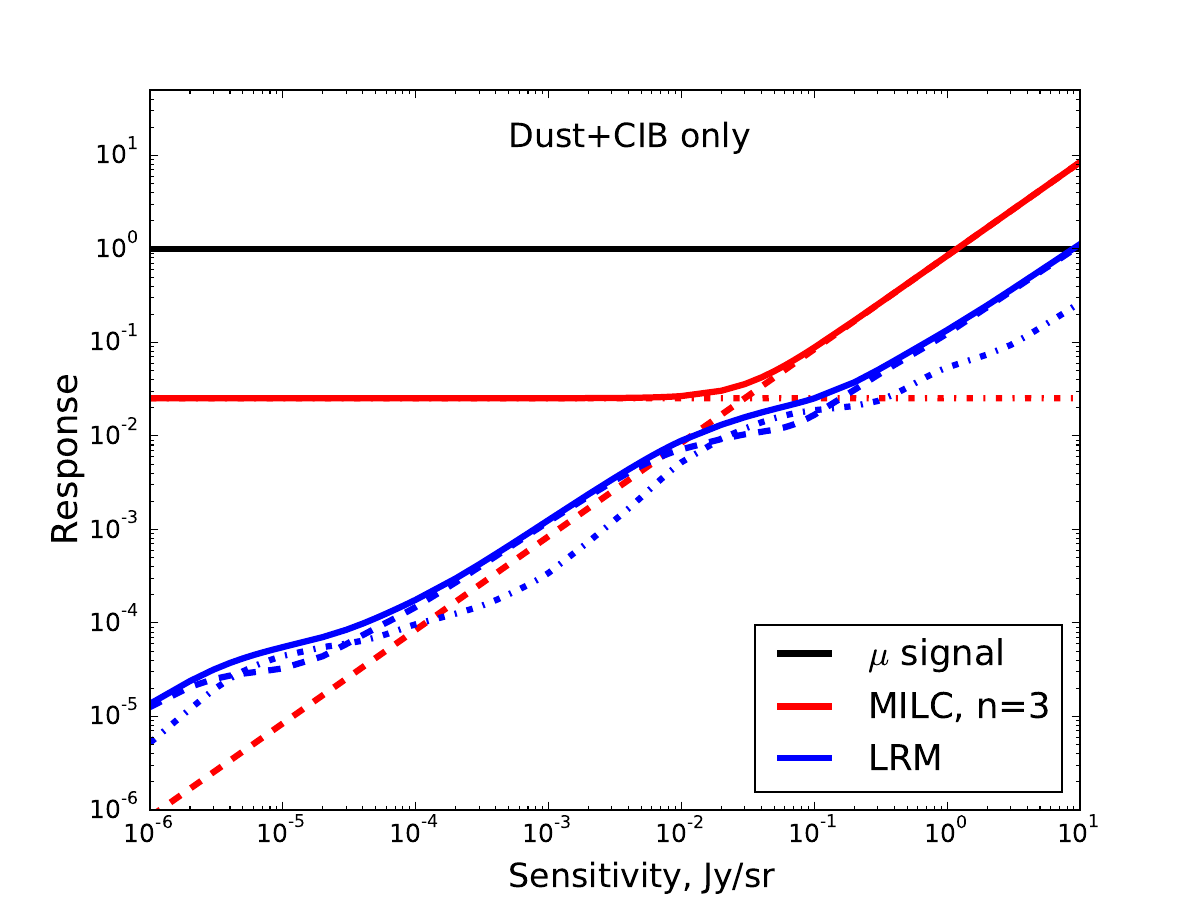}
  \caption{{\it Left panel}: red solid line: total response to
    foreground+noise for MILC as a function of the number of constraints.
    Red dashed line: Noise response for MILC. Red dash-dotted line:
    MILC response to foreground. The blue solid, dashed, and dash-dotted
    lines show LRM responses to foreground+noise, noise, and foreground,
    respectively. {\it Right panel}: responses to noise and foreground as
    functions of sensitivity for MILC (n=3) and LRM approaches. Solid,
    dashed, and dash-dotted lines represent the same as in the left panel.}
\end{figure*}

LRM, in contrast to MILC, assumes one single condition on foregrounds and
noise and provides us with the optimal weight vector $\boldsymbol{\omega}$
to minimize the ratio (foreground+noise)/signal.
The result can be seen in Fig. 1(h) and Fig. 1(i) for foreground and
foreground+noise responses, respectively and in Fig. 2.

An interesting approach is to study the responses to the
foreground and random noise on the sensitivity of the experiment:
$R({\bf F},\sigma)$ and $R({\bf N},\sigma)$.
This makes it possible to estimate the necessary sensitivity for reliable
detection of certain spectral distortions in the presence of foregrounds.
In addition, this dependence allows one to compare the effectiveness of
different methods of data cleaning.

Figure 2 (right panel) shows the response to foreground and noise as a
function of $\sigma$ for the MILC and LRM methods.
It is important to note that for the MILC method, the weight vector
${\boldsymbol\omega}$ does not depend on the sensitivity.
Thus, the dependence of the response to noise on $\sigma$ is linear
$\langle R^2({\bf N})\rangle^{{}^{\frac{1}{2}}}=\sigma
(\boldsymbol{\omega}\cdot\boldsymbol{\omega}^{{}^T})^{{}^{\frac{1}{2}}}$.
Since the response to the foreground is constant and given by Eq. (13),
the overall response to noise and signal as a function of
$\sigma$ has a simple analytical form:
\begin{equation}
  R_{_{MILC}}({\bf F}+{\bf N},\sigma)=
  \left[(\boldsymbol{\omega}\cdot\boldsymbol{\omega}^T)
  \sigma^2+R_{n+1}^2({\bf F})\right]^{\frac{1}{2}}.
\end{equation}  
Unlike the MILC method, when applying the LRM filtering, the weights
$\omega_j$ depend on the magnitude of the photon noise:
$\boldsymbol{\omega}=\boldsymbol{\omega}(\sigma)$.
As can be seen from Fig. 2, the total response to noise and foreground for
the LRM approach is always significantly less than for the MILS 
and provides a better signal/(foreground+noise) ratio:
\begin{equation}
 R_{_{LRM}}({\bf F}+{\bf N},\sigma)<R_{_{MILC}}({\bf F}+{\bf N},\sigma).
\end{equation}
This inequality is true for any sensitivity $\sigma$. The reason for this
is that LRM, unlike MILC, does not require complete orthogonality of the
investigated signal to the foreground.

\subsubsection{All foregrounds}

Below we perform an analysis of two approaches to the problem
of signal separation, taking into account all the main
foregrounds, adding them to the
already considered components of dust and infrared radiation.
As mentioned above, the signals of relic origin and free-free
emission are considered as well-defined components of the
spectrum without parameter variations. They are only subject to
restrictions from above in terms of amplitudes.

The model of synchrotron radiation depends on one parameter
$\beta_s$ which varies from 0.9 to 1.4. Empirically, it was
found that three derivatives of the expansion in a Taylor series in the vicinity of the reference point $\beta_s=1.15$
are sufficient to get rid of the response to this component for
MILC method. Thus, the synchrotron requires four constraints for the
weights $\boldsymbol{\omega}$ (the function itself and three
derivatives).

The spectrum emitted by the optical system of the instrument in
our example depends only on its temperature 
in the vicinity of $T_{opt}=10$ K and can vary between 9 K and 11 K.
To reliably get rid of the influence
of this radiation on data analysis, it is necessary to zero the
response to the function $B(\nu,T_{opt})$ and seven derivatives over
temperature: $\partial^n B(\nu,T_{opt})/\partial 
T_{opt}^n\mid_{T_{opt}=10K}$, 
$n\le 7$ (eight constraints in total).

\begin{figure*}[!htbp]
  \includegraphics[width=0.49\textwidth]{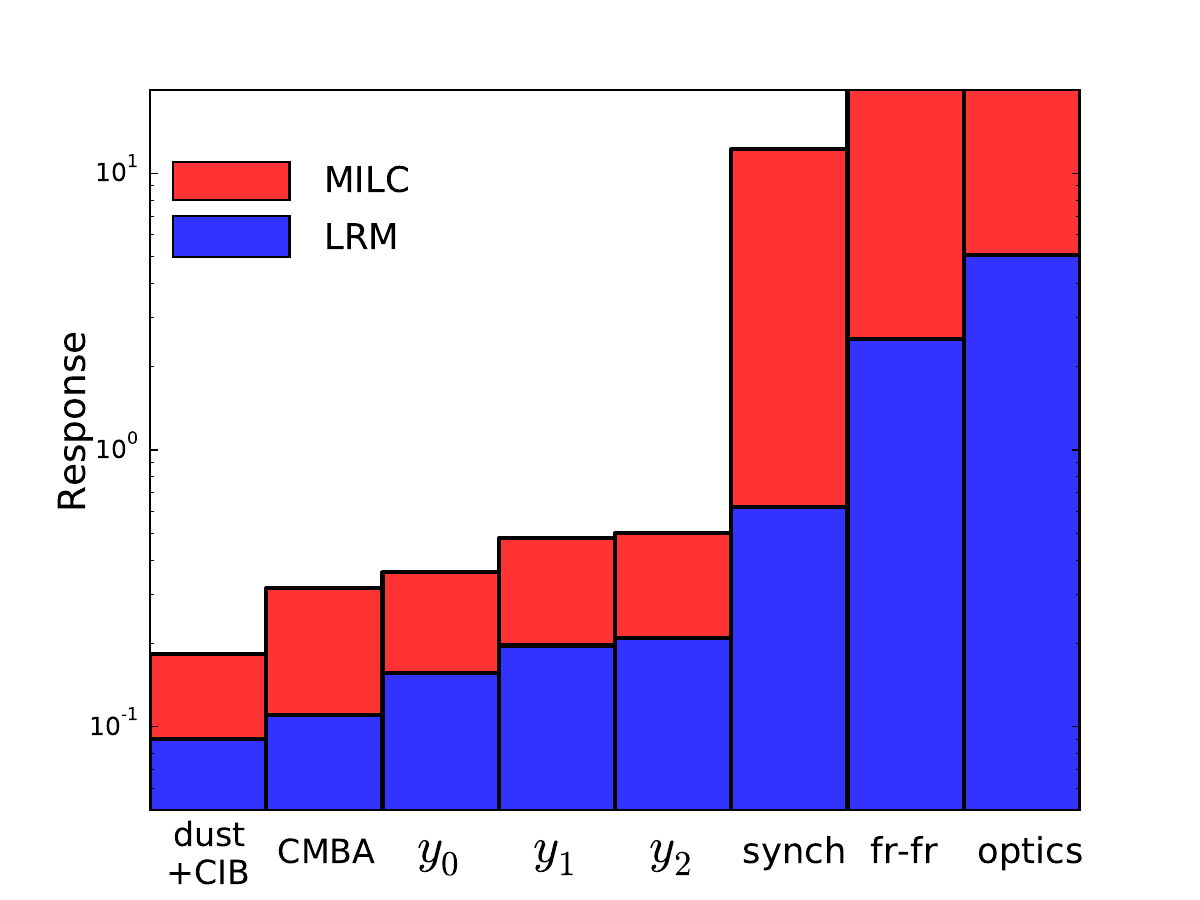}
  \includegraphics[width=0.49\textwidth]{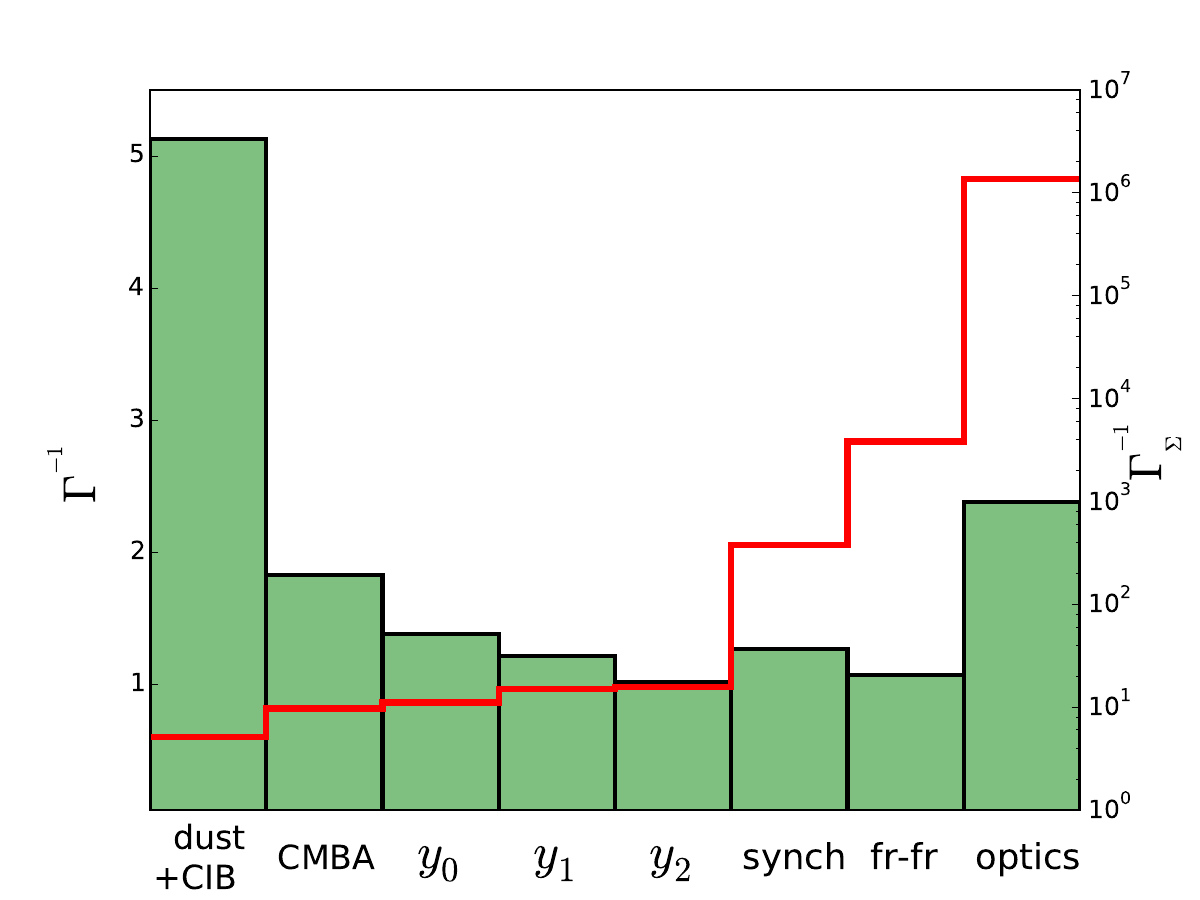}
  \caption{{\it Left panel}: the histogram represents the response to
    foreground+noise with the sequential addition
    (one by one from left to right) of various components
    to the studied signal. Thus, the leftmost column shows the response
    to foreground+noise in the presence of only dust and CIB radiation as
    foreground, and the rightmost column shows this response if
    all the components listed along the horizontal axis are taken into account.
    Calculations were made for MILC and LRM to separate $\mu$ distortions
    and correspond to
    the sensitivity $\sigma=1$ Jy/sr. {\it Right panel}: the green histogram
    shows the measure of orthogonality $\Gamma_c$ of the $\mu$ signal
    to each individual component. The red step line is the measure of
    orthogonality $\Gamma_\Sigma^{}$ of $\mu$ distortion to all components to
    the left of the step in question (similar to the left panel).
  }
\end{figure*}
\begin{figure*}[!htbp]
  \includegraphics[width=0.49\textwidth]{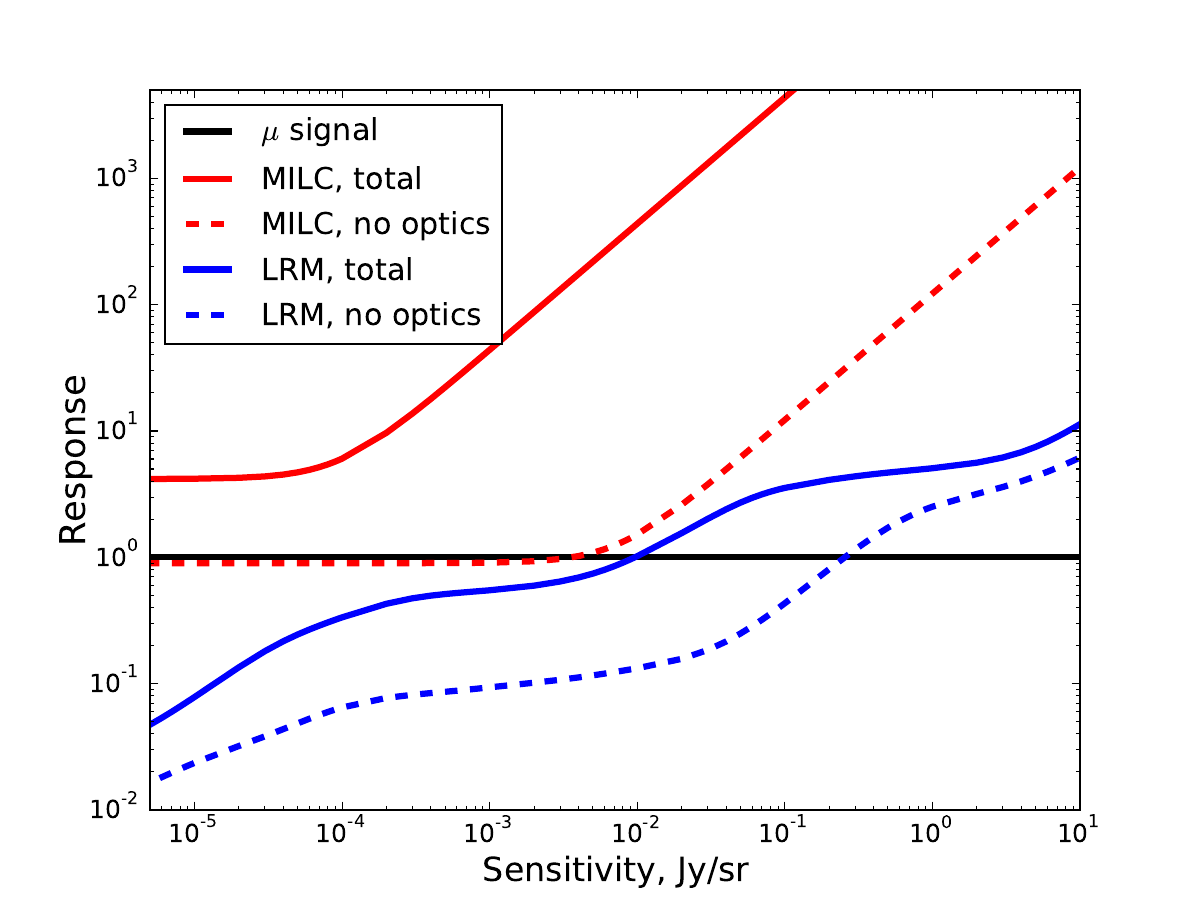}
  \includegraphics[width=0.49\textwidth]{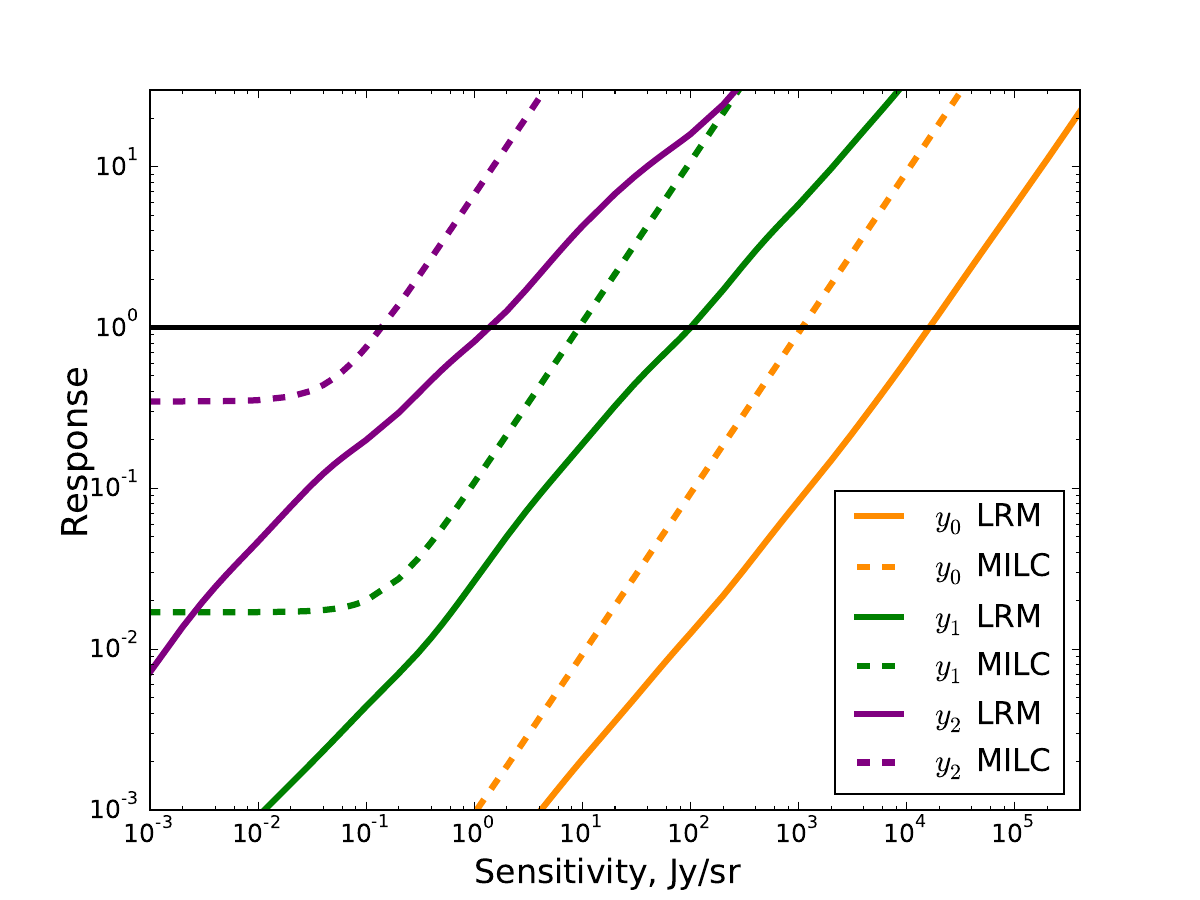}
  \caption {{\it Left panel}: the result of applying the MILC and LRM methods
    to separate the $\mu$ signal. The dependencies of the total
    noise+foreground response on the sensitivity of the experiment in the
    presence of all the main foregrounds are shown. The dashed lines show
    results for an 'ideal' experiment with no influence from the instrument
    optical system. {\it Right panel} shows the results of extracting
    $y_0$, $y_1$ and $y_2$ signals for the LRM and for MILC approaches. We show the
    responses to noise+foreground when taking into account all the main
    foregrounds except the instrument optics.}
\end{figure*}

The results for $\mu$ signal extraction for both methods are shown in Fig. 3
(left panel). The more components are taken into account, the larger the
foreground+noise response becomes for both methods. Nevertheless
LRM always shows an acceptable result, while MILC gives almost
infinitely large response to noise when trying to get rid of all foregrounds.
In the same figure (right panel) we show the
orthogonality measures of individual components to the $\mu$ signal:
$\Gamma_m=\Gamma_{dust+CIB}^{}, \Gamma_{{}_{CMBA}},...,\Gamma_{optics}$ along with a
orthogonality measure of the aggregate of components $\Gamma_\Sigma^{}$.
It is important to note that there is no direct analytical relation
between $\Gamma_{\Sigma}^{}$ and $\Gamma_m$ since the foreground signals are
not orthogonal to each other. It is easy to see,
that $R_{MILC}^{}({\bf F}+{\bf N})\sim\Gamma_{\Sigma}^{-1} $ as it should be.

Finally, Fig. 4 shows the dependence of the response
$R({\bf F}+{\bf N},\sigma)$ on the sensitivity of the experiment in the presence of all foregrounds for both methods for $\mu$ signal detection (left panel).
The right panel shows the same functions for the LRM approach in cases
when the signals of interest are $I_{y_0}, I_{y_1}, I_{y_2}$. In this case,
the foreground created by the instrument
is not taken into account, since when observing
$y$ distortions, it is possible to use the sky difference, which automatically
excludes this component from the observed signal.

\subsubsection{Optimal temperature for an instrument's optical system}

For most space experiments, it is preferable to cool the telescope's
primary mirror and other mirrors as much as possible to avoid creating
additional photon
noise and degrading the sensitivity. Below we will demonstrate that
cooling the system of mirrors to too much low temperature, that is, close to the
CMB temperature $\sim 3$ K, can significantly worsen the result if the
signal of interest is $\mu$-type distortion. Indeed, on the one hand,
a decrease in the instrument optics temperature leads to a decrease in
photon noise, on the other hand, the approach of this temperature to the
temperature of CMB reduces the degree of orthogonality of the $\mu$ signal to
the foreground created by the optics. Despite the fact that LRM does not
assume complete orthogonality to the foregrounds
\begin{figure}[!htbp]
  \includegraphics[width=1.0\columnwidth]{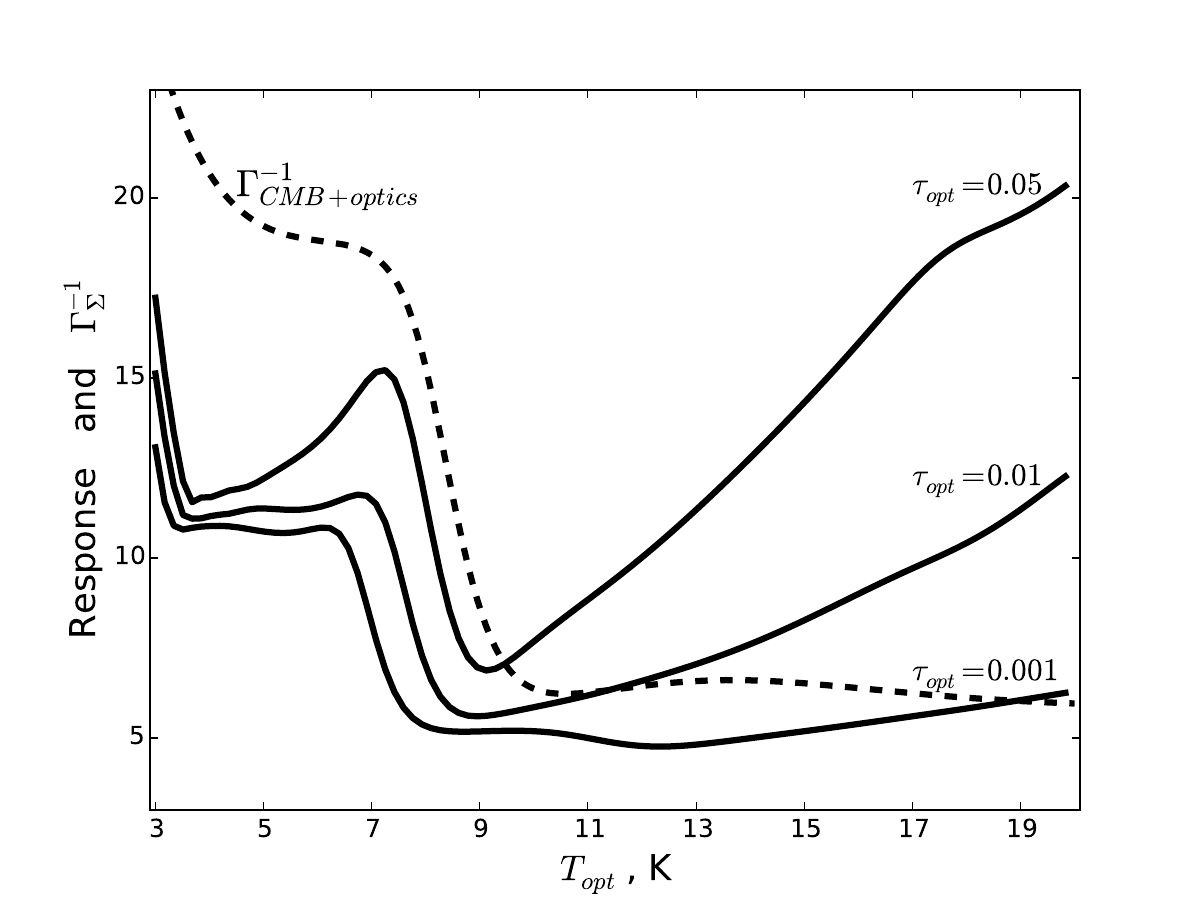}
  \caption {Solid lines show the response to foreground+noise as a function
    of the temperature of the instrument optics for different emissivity
    values. The dashed line corresponds to the value which is inverse to
    the measure of orthogonality between $I_{\mu}$ and a set of signals
    associated with CMB and optics:
    $I_{CMBA}^{}, I_{y_0}, I_{y_1}, I_{y_2}, I_{optics}$.}
\end{figure}
(including orthogonality to the signal created by the optics),
the combination
of CMB related signals together with a poorly defined signal from the optics
can, to a certain extent, mimic the desired $\mu$ signal. This inevitably
increases the response to foreground+noise and, thereby, reduces the
sensitivity of the experiment to $\mu$ distortion measurements. Figure 5
demonstrates this effect. Shown here are the responses
$R({\bf F}+{\bf N}, T_{opt})$ as a functions of the temperature of the optics for
three different values of optics emissivity $\tau_{opt}$ .
The responses are normalized in such a way that for all three considered $\tau_{opt}$,
a sensitivity of 1 $Jy/sr$ is achieved at $T_{opt}=10$ K. In the same figure we show
the orthogonality measure of the $I_\mu$ to the combination
of CMB related signals and optics: $I_{CMBA}^{}, I_{y_0}, I_{y_1}, I_{y_2}, I_{opt}$.
We denote this measure as $\Gamma_{CMB+optics}$.
It is assumed that the temperature can vary within $T_{opt}\pm 1$ K.
At low temperatures, this response behaves in a similar way as
$\Gamma_{CMB+optics}^{-1}$ and, therefore,
increases strongly if $T_{opt}\rightarrow T_{{}_{CMB}}$. At high temperatures the response
increases due to an increase in photon noise. This effect obviously
depends on the emissivity of the optics $\tau_{opt}$. Nevertheless, even
for small emissivity, cooling the mirror to temperatures close to $T_{{}_{CMB}}$
is undesirable. Finally the minimum response is reached when $T_{opt}\sim 9$ K.

It should be noted that the dependence we calculated will change for a
different FTS configuration, but the effect of increasing the response to
foreground+noise at $T_{opt}\sim T_{{}_{CMB}}$ will remain relatively strong
in any case.

\subsection{The instrument}

Below we present the example of a possible space mission for measuring CMB
spectral
distortions of $\mu$ type.
Figure 6 shows the top view of the instrument called SIMBAD (Spectroscopic
Interferometer for Microwave BAckground Distortions). The concept, originally
developed to be placed at L2, has been then proposed to be deposited in a
permanently shadowed region (PSR) present in some polar lunar craters
\citep{doi:10.1098/rsta.2020.0212}
since they could offer a very low temperature, passive cooling for the instrument.
The property of a few spots within the PSRs which reached $18$ K has been put
forward \citep{doi:10.1098/rsta.2023.0070}, as providing a much lower temperature than
is obtained at L2 for
JWST \citep{2006SSRv..123..485G} to the price of a huge sunscreen. The paper
\citep{workinprogress} is part of a special issue
devoted to the ``Astronomy from the Moon: the next decades''. In this version of SIMBAD (Fig. 6) the diameter of the telescope primary mirror is
pushed to 1.5 m, feeding a dual-input imaging FTS \citep{2013ExA....35..527M}.

All the instrument is placed in a cryostat designed to fit into the internal
fairing of the largest cargo launcher in development to go to the Moon
\citep{nasa}. To bend the beam, the telescope primary mirror and the following flat mirror must be actively cooled at $9$ K according to our results (Fig. 5). But the IFTS behind (right part of the cryostat) is actively cooled at a lower temperature, close to the CMB temperature, since its contribution to the foregrounds is cancelled by the differential concept of the instrument, reducing the photon noise budget. If the SIMBAD
cryostat is installed on one of the cold lunar spots at 18 K, as
described in \citep{doi:10.1098/rsta.2020.0212}, then such a low
temperature passive cooling could reduce the power needed for the active
cooling of the whole instrument.

At the two IFTS outputs there are $2\times 2$
bolometers with their feeding
horn as shown on Fig. 6, each of $1.5^{\circ}$ aperture on the sky,
providing a
total field coverage of $3^{\circ}\times 3^{\circ}$ for each data acquisition.
The spectral coverage of the IFTS is going from 15 to 2895 GHz, to cover the
full CMB domain. The maximum
motion of the moving mirror of the IFTS (6 mm) is adjusted to reach a 7.5 GHz
resolution on all the frequency domain, providing 384 equal frequency
channels. To be as simple as possible, the telescope has no tracking
motion. The direction
of pointing is determined by three adjustable legs which support the SIMBAD
cryostat
on the lunar ground. The scanning of a full ring of $3^{\circ}$ width
is simply
obtained by the rotation of the Moon. Thanks to the
IFTS \citep{2013ExA....35..527M}, the four bolometers are integrating
simultaneously, making a gain of a factor 2 in
sensitivity for $\mu$ compared to an FTS with a single detector. 
In paper \citep{doi:10.1098/rsta.2020.0212} the sensitivity of 1 Jy/sr is
reached by a SIMBAD instrument with a frequency resolution of 15 GHz and a
72 cm telescope diameter, but already a total field-of-view of
$3^{\circ}\times 3^{\circ}$ obtained by four detectors observing simultaneously
for
a total
integration time of 4 years.
\begin{figure}[!htbp]
  \includegraphics[width=1.0\columnwidth]{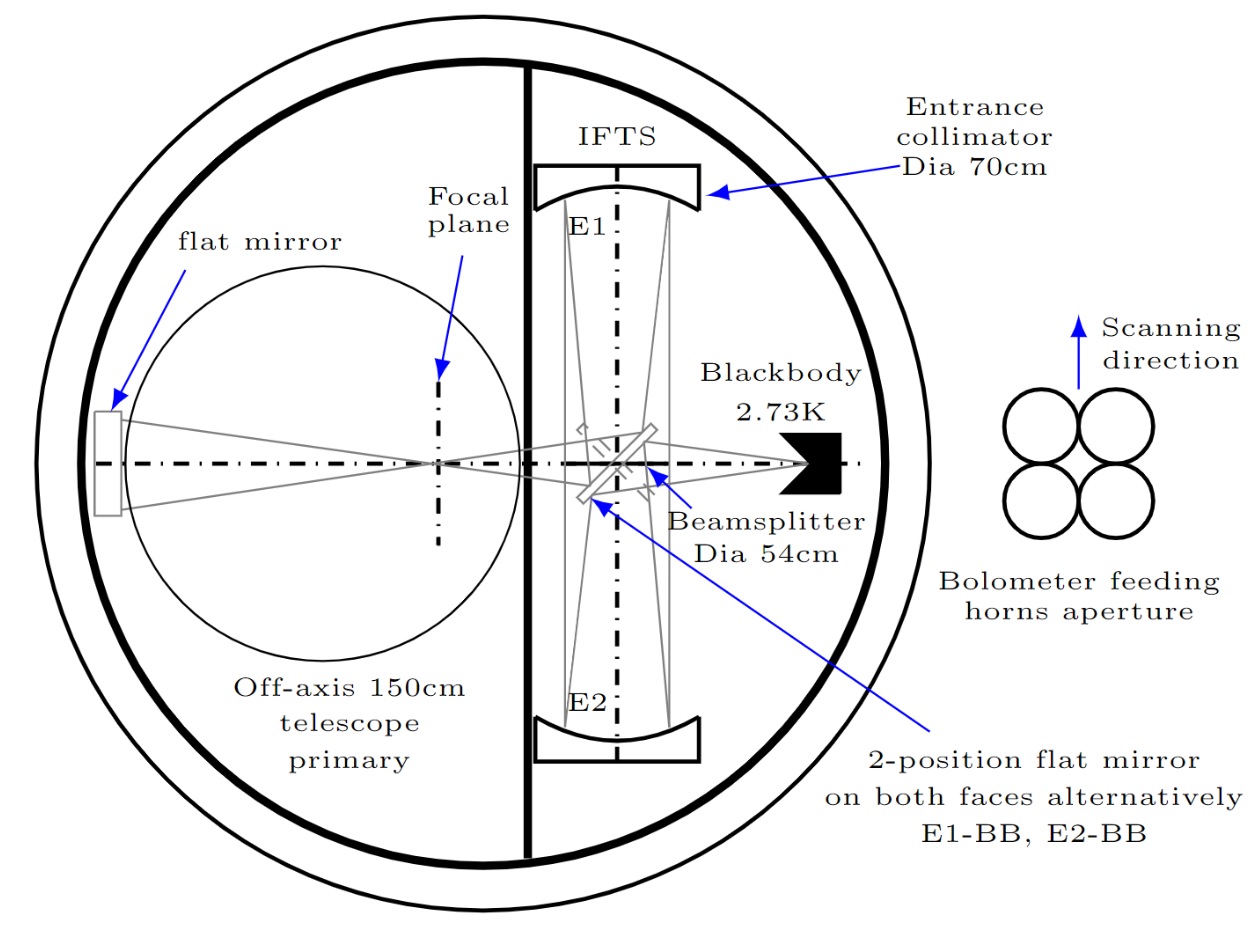}
  \caption{SIMBAD instrument}
\end{figure}
With the new SIMBAD with a frequency resolution
divided by 2 and a collecting area multiplied by 4, from Eq. (29), the same
total integration time is needed.
As proposed in \citep{doi:10.1098/rsta.2020.0212} the sensitivity and the
field of view can be improved by placing in the same lunar crater several identical SIMBAD units pointing on
adjacent directions scanning a wider ring. 

It is important to note that the instrument requirements for measuring
$\mu$, $y$ signals and recombination lines vary. For a $\mu$ signal,
it is not
important for us to have good angular resolution, but we need reliable
calibration of the instrument and the temperature of its optical system
should be about $9$ K. At the same time, y distortions and especially
relativistic corrections $y_1$, $y_2$ have a small angular size. For their
observations, it is desirable to have good angular resolution,
but there is no need to have a contrast between
the CMB temperature and the temperature of the instrument. Besides the
calibration of the instrument for this task is not required. To detect
extremely weak signals from hydrogen and helium recombination lines, it
is desirable to cool the instrument optics as much as possible to reduce
photon noise.

Thus, it is necessary to look for a very complex compromise to combine all
these requirements in one device if the goal of the mission is to
simultaneously solve all three problems. SIMBAD is mostly targeted
for $\mu$ signal measurements.

\section{Conclusions and Discussion}

Detecting CMB spectral distortions is one of the key goals of modern
observational cosmology. Hidden in foregrounds of cosmic and instrumental
origin, these deviations of the CMB spectrum from a perfect black body shape 
contain a wealth of information unobtainable by other observational methods.
To extract CMB frequency spectral distortions, a signal-foreground
separation algorithm was introduced. The key feature of this algorithm is
its weak sensitivity to the foregrounds spectral shapes. 

Given the areas of foreground parameters variation, which are relatively easy
to determine, and the upper bounds of the foreground amplitudes, whose
overestimation would not skew our results, we are able
to minimize the response both to the foregrounds and to noise. With this in
mind, we have compared the performance of our LRM algorithm with the MILC
method.
 
Starting off with a limited number of foregrounds (dust and CIB) we define
the minimal number of moments/constraints optimal for the MILC analysis.
Since LRM allows for the investigated signal to be nonorthogonal to the
foreground, the total response to noise and foreground for our method is
significantly less than for the MILC for any sensitivity value. Subsequently,
we broaden the scope of our analysis by including CMB anisotropy,
y-distortions, synchrotron and free-free emission and radiation created by
the instrument optics. Even though adding more components increases the total
response for both methods, the response to noise obtained with MILC exceeds
LRM results by a wide margin. In this paper we did not take into account Zodiacal light and CO lines as foregrounds in our modelling.
 
It is important to note that our LRM method allows for an improvement by
subtracting foreground components one by one from the observed signal.
Applying our method sequentially to all foreground components, one can
estimate the real foreground amplitudes $a_m$. By subtracting foregrounds
with estimated amplitudes from the total signal, new restrictions can be made
on the maximum possible amplitudes of foreground residuals $A_m$. Reducing
the upper limits of foreground amplitudes leads to a decrease in the response
to foreground+noise and, therefore, more accurate estimation of the spectral
distortion amplitudes. Implementation of such improvement is out of the
scope of this paper. 

Finally, we have established that cooling the instrument down to the temperature
close to the temperature of relic radiation spoils the results drastically
when measuring $\mu$ type distortions. This occurs mainly due to a decrease
in the measure of orthogonality of the $\mu$ signal to the instrument optics
component. If the system of mirrors is cooled to the temperature of the relic,
it itself begins to create distortions close in shape to the CMB spectral
distortions. The optimal result is reached when the temperature
of the mirrors is around $9$K.

These reasonings do not apply to the measurement of $y$ perturbations, since in
this case the signal difference can be used, which leads to the automatic
exclusion of the component created by the instrument optics from the
observational data. Also, these arguments are not relevant to the measurement
of recombination lines because their shape is not related to possible
distortions of the blackbody spectrum. Note that data processing for any physical
experiment with poorly defined foregrounds can use the
LRM approach.

\section*{ACKNOWLEDGMENT}

We wish to thank Jens Chluba for useful discussion.
The work is supported by the Project No. 41-2020
of LPI new scientific groups and the Foundation for the
Advancement of Theoretical Physics and Mathematics
BASIS, Grant No. 19-1-1-46-1.

\def\apj{Astrophys.~J}
\def\apjl{Astrophys.~J.,~Lett}
\def\apjs{Astrophys.~J.,~Supplement}
\def\an{Astron.~Nachr}     
\def\aap{Astron.~Astrophys}
\def\mnras{Mon.~Not.~R.~Astron.~Soc}
\def\pasp{Publ.~Astron.~Soc.~Pac}
\def\aaps{Astron.~and Astrophys.,~Suppl.~Ser}
\def\apss{Astrophys.~Space.~Sci}
\def\ibvs{Inf.~Bull.~Variable~Stars}
\def\japa{J.~Astrophys.~Astron}
\def\na{New~Astron}
\def\aspproc{Proc.~ASP~conf.~ser.}
\def\aspcs{ASP~Conf.~Ser}
\def\aj{Astron.~J}
\def\actaa{Acta Astron}
\def\araa{Ann.~Rev.~Astron.~Astrophys}
\def\caosp{Contrib.~Astron.~Obs.~Skalnat{\'e}~Pleso}
\def\pasj{Publ.~Astron.~Soc.~Jpn}
\def\memsai{Mem.~Soc.~Astron.~Ital}
\def\astl{Astron.~Letters}
\def\aipproc{Proc.~AIP~conf.~ser.}
\def\physrep{Physics Reports}
\def\jcap{Journal of Cosmology and Astroparticle Physics}
\def\baas{Bulletin of the AAS}
\def\ssr{Space~Sci.~Rev.}
\def\azh{Astronomicheskii Zhurnal}

\bibliography{a102.bib}




\end{document}